\documentclass[a4paper, onecolumn]{quantumarticle}
\pdfoutput=1
\usepackage{biblatex} 
\usepackage{graphicx} 
\usepackage{amsmath}
\usepackage{amssymb}
\usepackage{braket}
\usepackage{hyperref}
\usepackage{cleveref}
\usepackage{mathtools}
\usepackage{nicefrac}
\usepackage{placeins}
\usepackage{stmaryrd}
\usepackage{soul}
\usepackage{tikz}
\usepackage{xspace}
\usepackage{quantikz}

\usepackage{mathtools,xparse}

\usepackage{geometry}
\usepackage{placeins}

\usepackage{xcolor}
\definecolor{HighlightRed}{rgb}{0.9, 0.0, 0.0}

\definecolor{HighlightOrange}{RGB}{247, 146, 5}

\newcommand{\I}[1]{\mathcal{I}_{#1}}

\providecommand{\nkd}[3]{\left \llbracket #1,\, #2,\, #3 \right \rrbracket}

\title{Flagging the Clifford hierarchy:~Fault-tolerant logical $\frac{\pi}{2^l}$ rotations via measuring circuit gauge operators of non-Cliffords}
\author{Shival Dasu}
\affiliation{Quantinuum, Broomfield, CO 80021, USA}
\author{Ben Criger}
\affiliation{Quantinuum, Terrington House, Cambridge, CB2 1NL, UK}
\date{March 18, 2026}
\newtheorem{definition}{Definition}

\begin{document}

\maketitle

\begin{abstract}
  We provide a recursively defined sequence of flag circuits which will detect logical errors induced by non-fault-tolerant $R_{\overline{Z}}(\nicefrac{\pi}{2^l})$ gates on CSS codes with a fault distance of two. As applications, we give a family of circuits with $O(l)$ gates and ancillae which implement fault-tolerant $R_{\overline{Z}}(\frac{\pi}{2^l})$ or $R_{\overline{ZZ}}(\frac{\pi}{2^l})$ gates on any $\nkd{k + 2}{k}{2}$ iceberg code and fault-tolerant circuits of size $O(l)$ for preparing $\ket{\frac{\pi}{2^l}}$ resource states in the $\nkd{7}{1}{3}$ code, which can be used to perform fault-tolerant $R_{\overline{Z}}(\frac{\pi}{2^l})$ rotations via gate teleportation, allowing for implementations of these gates that bypass the high overheads of gate synthesis when $l$ is small relative to the precision required. We show how the circuits above can be generalized to $\pi( x_0.x_{1}x_{2}\ldots x_{l}) = \sum_{j=0}^{l}x_j\frac{\pi}{2^j}$ rotations with identical overheads in $l$, which could be useful in quantum simulations where time is digitized in binary. Finally, we illustrate two approaches to increase the fault-distance of our construction. We show how to increase the fault distance of a Cliffordized version of the T gate circuit to $3$ in the Steane code and how to increase the fault-distance of the $\nicefrac{\pi}{2}$ iceberg circuit to $4$ through concatenation in two-level iceberg codes. This yields a targeted logical $R_{\overline{Z}}(\nicefrac{\pi}{2})$ gate with fault distance $4$ on any row of logical qubits in an $\nkd{(k_2 + 2)(k_1+2)}{k_2k_1}{4}$ code.
\end{abstract}

\section{Introduction}

\begin{figure}[!htbp]
    \centering
    \begin{tikzpicture}[scale=1.000000,x=1pt,y=1pt]
\filldraw[color=white] (0.000000, -7.500000) rectangle (111.000000, 22.500000);
\draw[color=black] (0.000000,15.000000) -- (111.000000,15.000000);
\draw[color=black] (0.000000,15.000000) node[left] {$a$};
\draw[color=black] (0.000000,0.000000) -- (111.000000,0.000000);
\draw[color=black] (0.000000,0.000000) node[left] {$b$};
\begin{scope}
\draw[fill=white] (12.000000, 15.000000) +(-45.000000:8.485281pt and 8.485281pt) -- +(45.000000:8.485281pt and 8.485281pt) -- +(135.000000:8.485281pt and 8.485281pt) -- +(225.000000:8.485281pt and 8.485281pt) -- cycle;
\clip (12.000000, 15.000000) +(-45.000000:8.485281pt and 8.485281pt) -- +(45.000000:8.485281pt and 8.485281pt) -- +(135.000000:8.485281pt and 8.485281pt) -- +(225.000000:8.485281pt and 8.485281pt) -- cycle;
\draw (12.000000, 15.000000) node {$X$};
\end{scope}
\draw (33.000000,15.000000) -- (33.000000,0.000000);
\begin{scope}
\draw[fill=white] (33.000000, 0.000000) circle(3.000000pt);
\clip (33.000000, 0.000000) circle(3.000000pt);
\draw (30.000000, 0.000000) -- (36.000000, 0.000000);
\draw (33.000000, -3.000000) -- (33.000000, 3.000000);
\end{scope}
\filldraw (33.000000, 15.000000) circle(1.500000pt);
\begin{scope}
\draw[fill=white] (54.000000, 15.000000) +(-45.000000:8.485281pt and 8.485281pt) -- +(45.000000:8.485281pt and 8.485281pt) -- +(135.000000:8.485281pt and 8.485281pt) -- +(225.000000:8.485281pt and 8.485281pt) -- cycle;
\clip (54.000000, 15.000000) +(-45.000000:8.485281pt and 8.485281pt) -- +(45.000000:8.485281pt and 8.485281pt) -- +(135.000000:8.485281pt and 8.485281pt) -- +(225.000000:8.485281pt and 8.485281pt) -- cycle;
\draw (54.000000, 15.000000) node {$X$};
\end{scope}
\begin{scope}
\draw[fill=white] (54.000000, -0.000000) +(-45.000000:8.485281pt and 8.485281pt) -- +(45.000000:8.485281pt and 8.485281pt) -- +(135.000000:8.485281pt and 8.485281pt) -- +(225.000000:8.485281pt and 8.485281pt) -- cycle;
\clip (54.000000, -0.000000) +(-45.000000:8.485281pt and 8.485281pt) -- +(45.000000:8.485281pt and 8.485281pt) -- +(135.000000:8.485281pt and 8.485281pt) -- +(225.000000:8.485281pt and 8.485281pt) -- cycle;
\draw (54.000000, -0.000000) node {$X$};
\end{scope}
\draw[fill=white,color=white] (72.000000, -6.000000) rectangle (87.000000, 21.000000);
\draw (79.500000, 7.500000) node {$=$};
\draw (102.000000,15.000000) -- (102.000000,0.000000);
\begin{scope}
\draw[fill=white] (102.000000, 0.000000) circle(3.000000pt);
\clip (102.000000, 0.000000) circle(3.000000pt);
\draw (99.000000, 0.000000) -- (105.000000, 0.000000);
\draw (102.000000, -3.000000) -- (102.000000, 3.000000);
\end{scope}
\filldraw (102.000000, 15.000000) circle(1.500000pt);
\end{tikzpicture}
    \caption{A weight-three gauge operator of a CX gate.}
    \label{fig:cnot_xxx_gauge_operator}
\end{figure}
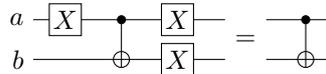

For any \emph{element} (state preparation, unitary gate, or measurement) within a quantum circuit, one can examine the gauge operators of that element which fix it. For Clifford circuits, it is natural to examine the Pauli operators at different spacetime locations which fix the circuit, which thereby generalizes the subsystem stabilizer code formalism to Clifford circuits \cite{BaconFlammiaHarrowShi, delfosse2023}. For example, supposing that the circuit in question consists of a $CX$ gate between two qubits $a$ and $b$, we can examine the local gauge group generators of $CX_{a,b}$. If we perform an $X_a$ operator before the $CX$ gate, this will cancel with an $X_a X_b$ operator after the $CX$ gate. In other words, $X_a$ before the $CX$ gate, followed by $X_aX_b$ after the $CX$ gate is a gauge operator of this gate, see Fig. \ref{fig:cnot_xxx_gauge_operator}.

\begin{figure}[!htbp]
    \centering
    \begin{tikzpicture}[scale=1.000000,x=1pt,y=1pt]
\filldraw[color=white] (0.000000, -7.500000) rectangle (102.000000, 22.500000);
\draw[color=black] (0.000000,15.000000) -- (102.000000,15.000000);
\draw[color=black] (0.000000,0.000000) -- (102.000000,0.000000);
\draw (27.000000,15.000000) -- (27.000000,0.000000);
\begin{scope}
\draw[fill=white] (27.000000, 7.500000) +(-45.000000:29.698485pt and 19.091883pt) -- +(45.000000:29.698485pt and 19.091883pt) -- +(135.000000:29.698485pt and 19.091883pt) -- +(225.000000:29.698485pt and 19.091883pt) -- cycle;
\clip (27.000000, 7.500000) +(-45.000000:29.698485pt and 19.091883pt) -- +(45.000000:29.698485pt and 19.091883pt) -- +(135.000000:29.698485pt and 19.091883pt) -- +(225.000000:29.698485pt and 19.091883pt) -- cycle;
\draw (27.000000, 7.500000) node {$U^{\dagger} V^{\dagger} U$};
\end{scope}
\draw (66.000000,15.000000) -- (66.000000,0.000000);
\begin{scope}
\draw[fill=white] (66.000000, 7.500000) +(-45.000000:8.485281pt and 19.091883pt) -- +(45.000000:8.485281pt and 19.091883pt) -- +(135.000000:8.485281pt and 19.091883pt) -- +(225.000000:8.485281pt and 19.091883pt) -- cycle;
\clip (66.000000, 7.500000) +(-45.000000:8.485281pt and 19.091883pt) -- +(45.000000:8.485281pt and 19.091883pt) -- +(135.000000:8.485281pt and 19.091883pt) -- +(225.000000:8.485281pt and 19.091883pt) -- cycle;
\draw (66.000000, 7.500000) node {$U$};
\end{scope}
\draw (90.000000,15.000000) -- (90.000000,0.000000);
\begin{scope}
\draw[fill=white] (90.000000, 7.500000) +(-45.000000:8.485281pt and 19.091883pt) -- +(45.000000:8.485281pt and 19.091883pt) -- +(135.000000:8.485281pt and 19.091883pt) -- +(225.000000:8.485281pt and 19.091883pt) -- cycle;
\clip (90.000000, 7.500000) +(-45.000000:8.485281pt and 19.091883pt) -- +(45.000000:8.485281pt and 19.091883pt) -- +(135.000000:8.485281pt and 19.091883pt) -- +(225.000000:8.485281pt and 19.091883pt) -- cycle;
\draw (90.000000, 7.500000) node {$V$};
\end{scope}
\end{tikzpicture}
    \caption{A gauge operator for an arbitrary unitary gate $U$, constructed by propagating an operator $V$ backwards through $U$.}
    \label{fig:arbitrary_U_gauge_operator}
\end{figure}
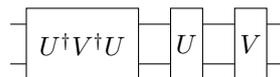

Given a circuit consisting of a single (possibly multi-qubit) unitary $U$, one can express the gauge operators of $U$ as in Fig. \ref{fig:arbitrary_U_gauge_operator}.

If $U$ is Clifford and $V$ is Pauli, then $UV^{\dagger}U^{\dagger}$ will be Pauli, naturally leading to the subsystem stabilizer code formalism of \cite{BaconFlammiaHarrowShi, delfosse2023}, and one can try to measure these operators to produce fault-tolerant circuits \cite{CrigerInPreparation}. But if $U$ is non-Clifford, this will not always be the case. Nevertheless, one can attempt to find circuits based on the measurement of these (possibly non-Pauli) gauge operators to upgrade the fault tolerance of an arbitrary $U$. We will call circuits that measure these operators flag circuits, in line with \cite{Chao_2018}.

In this work, we adapt this approach to $\nicefrac{\pi}{2^l}$ rotations. In particular, we introduce a recursive sequence of $l$ flag circuits which will detect logical errors induced by non-fault-tolerant $R_{\overline{Z}}(\frac{\pi}{2^l})$ rotations in CSS codes\footnote{In this work, we use the convention $R_P(\theta) \equiv \exp \left(-iP\nicefrac{\theta}{2}\right) $. }. We show that these flag circuits can be implemented to have a gate count, ancilla overhead, and depth of $O(l)$ and a fault distance of two, in the sense that a single gate error that induces a logical error will be detected by some measurement in a flag sub-circuit. Therefore, we first give this sequence in the case of the iceberg code \cite{gottesmanPHD, Self_2024}, yielding a family of circuits for any $\nkd{k+2}{k}{2}$ iceberg code $\mathcal{I}_k$, that will fault-tolerantly perform a logical $R_{\overline{Z}}(\frac{\pi}{2^l})$ gate on the $i$th logical qubit of $\I{k}$.

Also, the same circuits can be used to perform two-body $R_{\overline{Z}\overline{Z}}(\frac{\pi}{2^l})$ interactions between any two qubits $i$ and $j$ by changing the physical qubits of $\I{k}$ on which the circuit is supported, as well as $R_{\overline{X}}(\frac{\pi}{2^l})$, $R_{\overline{X}\overline{X}}(\frac{\pi}{2^l})$, $R_{\overline{Y}\overline{Y}}(\frac{\pi}{2^l})$, and $R_{\overline{X}/\overline{Y}/\overline{Z}^{\otimes k}}(\frac{\pi}{2^l})$, by permuting qubits and conjugating with transversal single-qubit gates. Also, we show how $R_{\overline{P}/\overline{PP}}(\pi( x_0.x_{1}x_{2}\ldots x_{l}) )$ rotations for any logical Pauli can be performed with the same overhead as logical $\nicefrac{\pi}{2^l}$ rotations, where $x_0.x_{1}x_{2}\ldots x_{l} = \sum_{j=0}^{l}x_j\frac{\pi}{2^j}$ is expressed in binary. Such rotations could be useful in quantum simulations where time is digitized to binary precision $l.$

Next, we discuss how this sequence can be generalized to non-fault-tolerant $R_{\overline{Z}}(\nicefrac{\pi}{2^l})$ rotations on other CSS codes. We illustrate this in the case of the $\nkd{7}{1}{3}$ code, giving a family of circuits which will produce $\ket{{\frac{\pi}{2^k}}}$ states in the Steane code with failure probability $O(p^2)$.

By implementing these exact angles using an approach based on flag circuits, we hope to obtain favorable overheads when compared to Clifford + $T$ gate synthesis in the case where a high degree of precision on a $\frac{\pi}{2^l}$ rotation is desired, such as in the examples above, since these circuits have overheads $O(l) = O(\log(\theta^{-1}))$ independent of accuracy, whereas Clifford+$T$ synthesis approaches have overheads of size $O(\log(\epsilon^{-1}))$, but apply to general single-qubit unitary operators. For this reason, this tradeoff also applies to approaches based on catalysis \cite{kim2026, Gidney_2018} or to angles which would otherwise require gate synthesis \cite{xu2026controlledjumpcliffordhierarchy}. For example, we find in state-vector simulations of the $\ket{\frac{\pi}{8}}$ state in the Steane code (albeit with a simplistic noise model and benchmarking) a logical infidelity of $\sim 4\times 10^{-6}$ produced with a circuit using only $5$ physical ancilla qubits and an acceptance rate of $86\%$, whereas \texttt{gridsynth}\cite{ross2016} requires $15$ magic states to be consumed in order to synthesize the rotation to sub-$10^{-3}$ infidelity. The fidelities and overheads of the $[[7,1,3]]$ circuit also suggests that the first non-trivial level of this construction could be used in conjunction with Clifford+$\sqrt{T}$ synthesis, which offers improved overheads relative to Clifford+$T$ \cite{Kliuchnikov_2023}.

Finally, we investigate increasing the fault distance of our construction through repeated gauge operator measurement in the Steane code, and concatenation in the iceberg code, in the case that the rotation angle is $\nicefrac{\pi}{2}$, so that the resulting circuit is Clifford and thus amenable to stabilizer simulation. In the former case, through measuring the main gauge operator of the rotation twice interleaved with two rounds of $Z$ syndrome extraction, we obtain a circuit with a fault-distance of 3. In the latter, we concatenate the $d=2$ $R_{\overline{Z}}(\nicefrac{\pi}{2})$ circuit with itself to obtain a targeted logical $R_{\overline{Z}}(\nicefrac{\pi}{2})$ gate with fault distance $4$ on a row of logical qubits in an $\nkd{(k_2 + 2)(k_1 + 2)}{k_2 k_1}{4}$ code.

\section{Fault-tolerant $R_{\overline{Z}}(\frac{\pi}{2^l})$ rotation circuits in the iceberg code}\label{sec:d=2 iceberg}

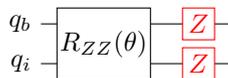
\begin{figure}[!htbp]
    \centering
    \begin{tikzpicture}[scale=1.000000,x=1pt,y=1pt]
\filldraw[color=white] (0.000000, -7.500000) rectangle (71.000000, 22.500000);
\draw[color=black] (0.000000,15.000000) -- (71.000000,15.000000);
\draw[color=black] (0.000000,15.000000) node[left] {$q_b$};
\draw[color=black] (0.000000,0.000000) -- (71.000000,0.000000);
\draw[color=black] (0.000000,0.000000) node[left] {$q_i$};
\draw (23.500000,15.000000) -- (23.500000,0.000000);
\begin{scope}
\draw[fill=white] (23.500000, 7.500000) +(-45.000000:24.748737pt and 19.091883pt) -- +(45.000000:24.748737pt and 19.091883pt) -- +(135.000000:24.748737pt and 19.091883pt) -- +(225.000000:24.748737pt and 19.091883pt) -- cycle;
\clip (23.500000, 7.500000) +(-45.000000:24.748737pt and 19.091883pt) -- +(45.000000:24.748737pt and 19.091883pt) -- +(135.000000:24.748737pt and 19.091883pt) -- +(225.000000:24.748737pt and 19.091883pt) -- cycle;
\draw (23.500000, 7.500000) node {$R_{ZZ}(\theta)$};
\end{scope}
\begin{scope}[color=red]
\begin{scope}[color=red]
\begin{scope}
\draw[fill=white] (59.000000, 15.000000) +(-45.000000:8.485281pt and 8.485281pt) -- +(45.000000:8.485281pt and 8.485281pt) -- +(135.000000:8.485281pt and 8.485281pt) -- +(225.000000:8.485281pt and 8.485281pt) -- cycle;
\clip (59.000000, 15.000000) +(-45.000000:8.485281pt and 8.485281pt) -- +(45.000000:8.485281pt and 8.485281pt) -- +(135.000000:8.485281pt and 8.485281pt) -- +(225.000000:8.485281pt and 8.485281pt) -- cycle;
\draw (59.000000, 15.000000) node {$Z$};
\end{scope}
\end{scope}
\end{scope}
\begin{scope}[color=red]
\begin{scope}[color=red]
\begin{scope}
\draw[fill=white] (59.000000, -0.000000) +(-45.000000:8.485281pt and 8.485281pt) -- +(45.000000:8.485281pt and 8.485281pt) -- +(135.000000:8.485281pt and 8.485281pt) -- +(225.000000:8.485281pt and 8.485281pt) -- cycle;
\clip (59.000000, -0.000000) +(-45.000000:8.485281pt and 8.485281pt) -- +(45.000000:8.485281pt and 8.485281pt) -- +(135.000000:8.485281pt and 8.485281pt) -- +(225.000000:8.485281pt and 8.485281pt) -- cycle;
\draw (59.000000, -0.000000) node {$Z$};
\end{scope}
\end{scope}
\end{scope}
\end{tikzpicture}
    \caption{Single gate failure of $ZZ$ type in a non-fault-tolerant $R_{ZZ_{q_i,q_b}}(\theta)$ rotation resulting in a logical $Z$ error.}
    \label{fig:non ft theta}
\end{figure}

The $\nkd{k + 2}{k}{2}$ iceberg code $\I{k}$ has logical $Z$ operators given by $\overline{Z}_i = Z_i Z_b$. We can perform a non-fault-tolerant  $R_{\overline{Z}}(\theta)$ by applying a single two-qubit gate, $R_{ZZ}(\theta)$, on physical qubits $q_i$ and $q_b$, since $R_{ZZ_{q_i, q_b}}(\theta) = e^{-i\frac{\theta}{2}\pi Z_i Z_b} = e^{-i\frac{\theta}{2} \pi \overline{Z}_i}$.
This is not fault-tolerant, because a single two-qubit gate failure can result in an $XX, YY$ or $ZZ$ error, see Fig. \ref{fig:non ft theta}.
All other errors are detectable: they cause a non-trivial syndrome when the $\I{k}$ stabilizers are measured. 

In the case of a maximally entangling Clifford gate, where $\theta = \frac{\pi}{2}$, we can make this circuit fault-tolerant by measuring Pauli gauge operators which fix this rotation and detect such errors\footnote{in order to detect a set of faults, a gauge operator must anticommute with the \emph{spackle} of the fault set, see \cite{BaconFlammiaHarrowShi, CrigerInPreparation}}. To detect $ZZ$ errors, we will measure a gauge operator formed by propagating an $X$ forward through the $R_{ZZ}\left( \nicefrac{\pi}{2} \right)$. To detect $XX$ and $YY$ errors which may be caused by the original $R_{ZZ}\left( \nicefrac{\pi}{2} \right)$, or by errors propagating through the flag sub-circuit, we measure an additional $ZZ$ gauge operator, see Fig. \ref{fig:ZZ_pi_by_2}.

\begin{figure}[!htbp]
    \centering
    \begin{tikzpicture}[scale=1.000000,x=1pt,y=1pt]
\filldraw[color=white] (0.000000, -7.500000) rectangle (95.000000, 22.500000);
\draw[color=black] (0.000000,15.000000) -- (95.000000,15.000000);
\draw[color=black] (0.000000,15.000000) node[left] {$q_b$};
\draw[color=black] (0.000000,0.000000) -- (95.000000,0.000000);
\draw[color=black] (0.000000,0.000000) node[left] {$q_i$};
\begin{scope}
\draw[fill=white] (12.000000, -0.000000) +(-45.000000:8.485281pt and 8.485281pt) -- +(45.000000:8.485281pt and 8.485281pt) -- +(135.000000:8.485281pt and 8.485281pt) -- +(225.000000:8.485281pt and 8.485281pt) -- cycle;
\clip (12.000000, -0.000000) +(-45.000000:8.485281pt and 8.485281pt) -- +(45.000000:8.485281pt and 8.485281pt) -- +(135.000000:8.485281pt and 8.485281pt) -- +(225.000000:8.485281pt and 8.485281pt) -- cycle;
\draw (12.000000, -0.000000) node {$X$};
\end{scope}
\draw (47.500000,15.000000) -- (47.500000,0.000000);
\begin{scope}
\draw[fill=white] (47.500000, 7.500000) +(-45.000000:24.748737pt and 19.091883pt) -- +(45.000000:24.748737pt and 19.091883pt) -- +(135.000000:24.748737pt and 19.091883pt) -- +(225.000000:24.748737pt and 19.091883pt) -- cycle;
\clip (47.500000, 7.500000) +(-45.000000:24.748737pt and 19.091883pt) -- +(45.000000:24.748737pt and 19.091883pt) -- +(135.000000:24.748737pt and 19.091883pt) -- +(225.000000:24.748737pt and 19.091883pt) -- cycle;
\draw (47.500000, 7.500000) node {$R_{ZZ}(\frac{\pi}{2})$};
\end{scope}
\begin{scope}
\draw[fill=white] (83.000000, 15.000000) +(-45.000000:8.485281pt and 8.485281pt) -- +(45.000000:8.485281pt and 8.485281pt) -- +(135.000000:8.485281pt and 8.485281pt) -- +(225.000000:8.485281pt and 8.485281pt) -- cycle;
\clip (83.000000, 15.000000) +(-45.000000:8.485281pt and 8.485281pt) -- +(45.000000:8.485281pt and 8.485281pt) -- +(135.000000:8.485281pt and 8.485281pt) -- +(225.000000:8.485281pt and 8.485281pt) -- cycle;
\draw (83.000000, 15.000000) node {$Z$};
\end{scope}
\begin{scope}
\draw[fill=white] (83.000000, -0.000000) +(-45.000000:8.485281pt and 8.485281pt) -- +(45.000000:8.485281pt and 8.485281pt) -- +(135.000000:8.485281pt and 8.485281pt) -- +(225.000000:8.485281pt and 8.485281pt) -- cycle;
\clip (83.000000, -0.000000) +(-45.000000:8.485281pt and 8.485281pt) -- +(45.000000:8.485281pt and 8.485281pt) -- +(135.000000:8.485281pt and 8.485281pt) -- +(225.000000:8.485281pt and 8.485281pt) -- cycle;
\draw (83.000000, -0.000000) node {$Y$};
\end{scope}
\end{tikzpicture} \hspace{0.1\textwidth} \begin{tikzpicture}[scale=1.000000,x=1pt,y=1pt]
\filldraw[color=white] (0.000000, -7.500000) rectangle (100.000000, 22.500000);
\draw[color=black] (0.000000,15.000000) -- (100.000000,15.000000);
\draw[color=black] (0.000000,15.000000) node[left] {$q_b$};
\draw[color=black] (0.000000,0.000000) -- (100.000000,0.000000);
\draw[color=black] (0.000000,0.000000) node[left] {$q_i$};
\begin{scope}
\draw[fill=white] (12.000000, 15.000000) +(-45.000000:8.485281pt and 8.485281pt) -- +(45.000000:8.485281pt and 8.485281pt) -- +(135.000000:8.485281pt and 8.485281pt) -- +(225.000000:8.485281pt and 8.485281pt) -- cycle;
\clip (12.000000, 15.000000) +(-45.000000:8.485281pt and 8.485281pt) -- +(45.000000:8.485281pt and 8.485281pt) -- +(135.000000:8.485281pt and 8.485281pt) -- +(225.000000:8.485281pt and 8.485281pt) -- cycle;
\draw (12.000000, 15.000000) node {$Z$};
\end{scope}
\draw (50.000000,15.000000) -- (50.000000,0.000000);
\begin{scope}
\draw[fill=white] (50.000000, 7.500000) +(-45.000000:28.284271pt and 19.091883pt) -- +(45.000000:28.284271pt and 19.091883pt) -- +(135.000000:28.284271pt and 19.091883pt) -- +(225.000000:28.284271pt and 19.091883pt) -- cycle;
\clip (50.000000, 7.500000) +(-45.000000:28.284271pt and 19.091883pt) -- +(45.000000:28.284271pt and 19.091883pt) -- +(135.000000:28.284271pt and 19.091883pt) -- +(225.000000:28.284271pt and 19.091883pt) -- cycle;
\draw (50.000000, 7.500000) node {$R_{ZZ}(\nicefrac{\pi}{2})$};
\end{scope}
\begin{scope}
\draw[fill=white] (88.000000, 15.000000) +(-45.000000:8.485281pt and 8.485281pt) -- +(45.000000:8.485281pt and 8.485281pt) -- +(135.000000:8.485281pt and 8.485281pt) -- +(225.000000:8.485281pt and 8.485281pt) -- cycle;
\clip (88.000000, 15.000000) +(-45.000000:8.485281pt and 8.485281pt) -- +(45.000000:8.485281pt and 8.485281pt) -- +(135.000000:8.485281pt and 8.485281pt) -- +(225.000000:8.485281pt and 8.485281pt) -- cycle;
\draw (88.000000, 15.000000) node {$Z$};
\end{scope}
\end{tikzpicture}\\
    \hspace{0.1cm}
    \begin{tikzpicture}[scale=1.000000,x=1pt,y=1pt]
\filldraw[color=white] (0.000000, -7.500000) rectangle (107.000000, 52.500000);
\draw[color=black] (0.000000,45.000000) -- (107.000000,45.000000);
\draw[color=black] (0.000000,45.000000) node[left] {$|0\rangle$};
\draw[color=black] (0.000000,30.000000) -- (107.000000,30.000000);
\draw[color=black] (0.000000,30.000000) node[left] {$q_b$};
\draw[color=black] (0.000000,15.000000) -- (107.000000,15.000000);
\draw[color=black] (0.000000,15.000000) node[left] {$q_i$};
\draw[color=black] (0.000000,0.000000) -- (107.000000,0.000000);
\draw[color=black] (0.000000,0.000000) node[left] {$|+\rangle$};
\draw (9.000000,45.000000) -- (9.000000,30.000000);
\begin{scope}
\draw[fill=white] (9.000000, 45.000000) circle(3.000000pt);
\clip (9.000000, 45.000000) circle(3.000000pt);
\draw (6.000000, 45.000000) -- (12.000000, 45.000000);
\draw (9.000000, 42.000000) -- (9.000000, 48.000000);
\end{scope}
\filldraw (9.000000, 30.000000) circle(1.500000pt);
\draw (9.000000,15.000000) -- (9.000000,0.000000);
\begin{scope}
\draw[fill=white] (9.000000, 15.000000) circle(3.000000pt);
\clip (9.000000, 15.000000) circle(3.000000pt);
\draw (6.000000, 15.000000) -- (12.000000, 15.000000);
\draw (9.000000, 12.000000) -- (9.000000, 18.000000);
\end{scope}
\filldraw (9.000000, 0.000000) circle(1.500000pt);
\draw (41.500000,30.000000) -- (41.500000,15.000000);
\begin{scope}
\draw[fill=white] (41.500000, 22.500000) +(-45.000000:24.748737pt and 19.091883pt) -- +(45.000000:24.748737pt and 19.091883pt) -- +(135.000000:24.748737pt and 19.091883pt) -- +(225.000000:24.748737pt and 19.091883pt) -- cycle;
\clip (41.500000, 22.500000) +(-45.000000:24.748737pt and 19.091883pt) -- +(45.000000:24.748737pt and 19.091883pt) -- +(135.000000:24.748737pt and 19.091883pt) -- +(225.000000:24.748737pt and 19.091883pt) -- cycle;
\draw (41.500000, 22.500000) node {$R_{ZZ}(\frac{\pi}{2})$};
\end{scope}
\draw (74.000000,30.000000) -- (74.000000,0.000000);
\filldraw (74.000000, 30.000000) circle(1.500000pt);
\filldraw (74.000000, 0.000000) circle(1.500000pt);
\draw (95.000000,15.000000) -- (95.000000,0.000000);
\begin{scope}
\draw[fill=white] (95.000000, 15.000000) +(-45.000000:8.485281pt and 8.485281pt) -- +(45.000000:8.485281pt and 8.485281pt) -- +(135.000000:8.485281pt and 8.485281pt) -- +(225.000000:8.485281pt and 8.485281pt) -- cycle;
\clip (95.000000, 15.000000) +(-45.000000:8.485281pt and 8.485281pt) -- +(45.000000:8.485281pt and 8.485281pt) -- +(135.000000:8.485281pt and 8.485281pt) -- +(225.000000:8.485281pt and 8.485281pt) -- cycle;
\draw (95.000000, 15.000000) node {$Y$};
\end{scope}
\filldraw (95.000000, 0.000000) circle(1.500000pt);
\draw (95.000000,45.000000) -- (95.000000,30.000000);
\begin{scope}
\draw[fill=white] (95.000000, 45.000000) circle(3.000000pt);
\clip (95.000000, 45.000000) circle(3.000000pt);
\draw (92.000000, 45.000000) -- (98.000000, 45.000000);
\draw (95.000000, 42.000000) -- (95.000000, 48.000000);
\end{scope}
\filldraw (95.000000, 30.000000) circle(1.500000pt);
\end{tikzpicture}
    \caption{Top: circuit gauge operators for $R_{ZZ}(\nicefrac{\pi}{2})$ that anticommute with $XX$, $YY$, or $ZZ$ errors occurring immediately after the $R_{ZZ}(\nicefrac{\pi}{2})$ gate. Bottom: A flag circuit which measures these gauge operators, achieving fault distance 2.}
    \label{fig:ZZ_pi_by_2}
\end{figure}

We call this circuit $\overline{R_{ZZ_{q_i,q_b}}(\nicefrac{\pi}{2})}^{FT}$ in the remainder of this work.
This Clifford circuit for a targeted logical phase gate (which can be modified to produce a global phase gate) on the iceberg code is not novel (see \cite{Chao_2018, Berthusen_2025}), but it illustrates the procedure we will use to recursively construct fault-tolerant non-Clifford circuits below.

This circuit is the first in the hierarchy of fault-tolerant $R_{ZZ}(\nicefrac{\pi}{2^l})$ circuits. Before deriving the entire family, we make some observations about the $R_{ZZ}(\nicefrac{\pi}{4})$ case (which will perform a logical $T$ gate) to motivate the recursion. We begin by observing that propagating an $X$ operator through an $R_{ZZ}(\theta)$ flips the sign of the rotation angle:
\begin{equation}
    \begin{tikzpicture}[scale=1.000000,x=1pt,y=1pt]
\filldraw[color=white] (0.000000, -7.500000) rectangle (174.000000, 22.500000);
\draw[color=black] (0.000000,15.000000) -- (174.000000,15.000000);
\draw[color=black] (0.000000,15.000000) node[left] {$q_b$};
\draw[color=black] (0.000000,0.000000) -- (174.000000,0.000000);
\draw[color=black] (0.000000,0.000000) node[left] {$q_i$};
\begin{scope}
\draw[fill=white] (12.000000, -0.000000) +(-45.000000:8.485281pt and 8.485281pt) -- +(45.000000:8.485281pt and 8.485281pt) -- +(135.000000:8.485281pt and 8.485281pt) -- +(225.000000:8.485281pt and 8.485281pt) -- cycle;
\clip (12.000000, -0.000000) +(-45.000000:8.485281pt and 8.485281pt) -- +(45.000000:8.485281pt and 8.485281pt) -- +(135.000000:8.485281pt and 8.485281pt) -- +(225.000000:8.485281pt and 8.485281pt) -- cycle;
\draw (12.000000, -0.000000) node {$X$};
\end{scope}
\draw (47.500000,15.000000) -- (47.500000,0.000000);
\begin{scope}
\draw[fill=white] (47.500000, 7.500000) +(-45.000000:24.748737pt and 19.091883pt) -- +(45.000000:24.748737pt and 19.091883pt) -- +(135.000000:24.748737pt and 19.091883pt) -- +(225.000000:24.748737pt and 19.091883pt) -- cycle;
\clip (47.500000, 7.500000) +(-45.000000:24.748737pt and 19.091883pt) -- +(45.000000:24.748737pt and 19.091883pt) -- +(135.000000:24.748737pt and 19.091883pt) -- +(225.000000:24.748737pt and 19.091883pt) -- cycle;
\draw (47.500000, 7.500000) node {$R_{ZZ}(\theta)$};
\end{scope}
\draw[fill=white,color=white] (77.000000, -6.000000) rectangle (92.000000, 21.000000);
\draw (84.500000, 7.500000) node {$=$};
\draw (124.000000,15.000000) -- (124.000000,0.000000);
\begin{scope}
\draw[fill=white] (124.000000, 7.500000) +(-45.000000:28.284271pt and 19.091883pt) -- +(45.000000:28.284271pt and 19.091883pt) -- +(135.000000:28.284271pt and 19.091883pt) -- +(225.000000:28.284271pt and 19.091883pt) -- cycle;
\clip (124.000000, 7.500000) +(-45.000000:28.284271pt and 19.091883pt) -- +(45.000000:28.284271pt and 19.091883pt) -- +(135.000000:28.284271pt and 19.091883pt) -- +(225.000000:28.284271pt and 19.091883pt) -- cycle;
\draw (124.000000, 7.500000) node {$R_{ZZ}(-\theta)$};
\end{scope}
\begin{scope}
\draw[fill=white] (162.000000, -0.000000) +(-45.000000:8.485281pt and 8.485281pt) -- +(45.000000:8.485281pt and 8.485281pt) -- +(135.000000:8.485281pt and 8.485281pt) -- +(225.000000:8.485281pt and 8.485281pt) -- cycle;
\clip (162.000000, -0.000000) +(-45.000000:8.485281pt and 8.485281pt) -- +(45.000000:8.485281pt and 8.485281pt) -- +(135.000000:8.485281pt and 8.485281pt) -- +(225.000000:8.485281pt and 8.485281pt) -- cycle;
\draw (162.000000, -0.000000) node {$X$};
\end{scope}
\end{tikzpicture}
\end{equation}.

Therefore, $R_{ZZ_{q_b,q_i}}(\theta)$ has the gauge operator below for all $\theta$:

\begin{equation}
    \begin{tikzpicture}[scale=1.000000,x=1pt,y=1pt]
\filldraw[color=white] (0.000000, -7.500000) rectangle (231.000000, 22.500000);
\draw[color=black] (0.000000,15.000000) -- (231.000000,15.000000);
\draw[color=black] (0.000000,15.000000) node[left] {$q_b$};
\draw[color=black] (0.000000,0.000000) -- (231.000000,0.000000);
\draw[color=black] (0.000000,0.000000) node[left] {$q_i$};
\begin{scope}
\draw[fill=white] (12.000000, -0.000000) +(-45.000000:8.485281pt and 8.485281pt) -- +(45.000000:8.485281pt and 8.485281pt) -- +(135.000000:8.485281pt and 8.485281pt) -- +(225.000000:8.485281pt and 8.485281pt) -- cycle;
\clip (12.000000, -0.000000) +(-45.000000:8.485281pt and 8.485281pt) -- +(45.000000:8.485281pt and 8.485281pt) -- +(135.000000:8.485281pt and 8.485281pt) -- +(225.000000:8.485281pt and 8.485281pt) -- cycle;
\draw (12.000000, -0.000000) node {$X$};
\end{scope}
\draw (47.500000,15.000000) -- (47.500000,0.000000);
\begin{scope}
\draw[fill=white] (47.500000, 7.500000) +(-45.000000:24.748737pt and 19.091883pt) -- +(45.000000:24.748737pt and 19.091883pt) -- +(135.000000:24.748737pt and 19.091883pt) -- +(225.000000:24.748737pt and 19.091883pt) -- cycle;
\clip (47.500000, 7.500000) +(-45.000000:24.748737pt and 19.091883pt) -- +(45.000000:24.748737pt and 19.091883pt) -- +(135.000000:24.748737pt and 19.091883pt) -- +(225.000000:24.748737pt and 19.091883pt) -- cycle;
\draw (47.500000, 7.500000) node {$R_{ZZ}(\theta)$};
\end{scope}
\draw (102.000000,15.000000) -- (102.000000,0.000000);
\begin{scope}
\draw[fill=white] (102.000000, 7.500000) +(-45.000000:35.355339pt and 19.091883pt) -- +(45.000000:35.355339pt and 19.091883pt) -- +(135.000000:35.355339pt and 19.091883pt) -- +(225.000000:35.355339pt and 19.091883pt) -- cycle;
\clip (102.000000, 7.500000) +(-45.000000:35.355339pt and 19.091883pt) -- +(45.000000:35.355339pt and 19.091883pt) -- +(135.000000:35.355339pt and 19.091883pt) -- +(225.000000:35.355339pt and 19.091883pt) -- cycle;
\draw (102.000000, 7.500000) node {$R_{ZZ}(-2\theta)$};
\end{scope}
\begin{scope}
\draw[fill=white] (145.000000, -0.000000) +(-45.000000:8.485281pt and 8.485281pt) -- +(45.000000:8.485281pt and 8.485281pt) -- +(135.000000:8.485281pt and 8.485281pt) -- +(225.000000:8.485281pt and 8.485281pt) -- cycle;
\clip (145.000000, -0.000000) +(-45.000000:8.485281pt and 8.485281pt) -- +(45.000000:8.485281pt and 8.485281pt) -- +(135.000000:8.485281pt and 8.485281pt) -- +(225.000000:8.485281pt and 8.485281pt) -- cycle;
\draw (145.000000, -0.000000) node {$X$};
\end{scope}
\draw[fill=white,color=white] (163.000000, -6.000000) rectangle (178.000000, 21.000000);
\draw (170.500000, 7.500000) node {$=$};
\draw (207.500000,15.000000) -- (207.500000,0.000000);
\begin{scope}
\draw[fill=white] (207.500000, 7.500000) +(-45.000000:24.748737pt and 19.091883pt) -- +(45.000000:24.748737pt and 19.091883pt) -- +(135.000000:24.748737pt and 19.091883pt) -- +(225.000000:24.748737pt and 19.091883pt) -- cycle;
\clip (207.500000, 7.500000) +(-45.000000:24.748737pt and 19.091883pt) -- +(45.000000:24.748737pt and 19.091883pt) -- +(135.000000:24.748737pt and 19.091883pt) -- +(225.000000:24.748737pt and 19.091883pt) -- cycle;
\draw (207.500000, 7.500000) node {$R_{ZZ}(\theta)$};
\end{scope}
\end{tikzpicture}
\end{equation}

Measuring this gauge operator using the flag qubit $a_0$ will detect a $ZZ$ error (shown in blue) coming from the $R_{ZZ}(\nicefrac{\pi}{4})$ gate in Fig.~\ref{fig:nonft_piby4}.
\begin{figure}[!htbp]
    \centering
    \begin{tikzpicture}[scale=1.000000,x=1pt,y=1pt]
\filldraw[color=white] (0.000000, -7.500000) rectangle (249.000000, 67.500000);
\draw[color=black] (0.000000,60.000000) -- (213.000000,60.000000);
\draw[color=black] (213.000000,59.500000) -- (249.000000,59.500000);
\draw[color=black] (213.000000,60.500000) -- (249.000000,60.500000);
\draw[color=black] (0.000000,60.000000) node[left] {$a_z\colon|0\rangle$};
\draw[color=black] (0.000000,45.000000) -- (249.000000,45.000000);
\draw[color=black] (0.000000,45.000000) node[left] {$q_b$};
\draw[color=black] (0.000000,30.000000) -- (249.000000,30.000000);
\draw[color=black] (0.000000,30.000000) node[left] {$q_i$};
\draw[color=black] (0.000000,15.000000) -- (237.000000,15.000000);
\draw[color=black] (237.000000,14.500000) -- (249.000000,14.500000);
\draw[color=black] (237.000000,15.500000) -- (249.000000,15.500000);
\draw[color=black] (0.000000,15.000000) node[left] {$a_0\colon|+\rangle$};
\draw[color=black] (0.000000,0.000000) -- (237.000000,0.000000);
\draw[color=black] (237.000000,-0.500000) -- (249.000000,-0.500000);
\draw[color=black] (237.000000,0.500000) -- (249.000000,0.500000);
\draw[color=black] (0.000000,0.000000) node[left] {$a_1\colon|0\rangle$};
\draw (9.000000,60.000000) -- (9.000000,45.000000);
\begin{scope}
\draw[fill=white] (9.000000, 60.000000) circle(3.000000pt);
\clip (9.000000, 60.000000) circle(3.000000pt);
\draw (6.000000, 60.000000) -- (12.000000, 60.000000);
\draw (9.000000, 57.000000) -- (9.000000, 63.000000);
\end{scope}
\filldraw (9.000000, 45.000000) circle(1.500000pt);
\draw (9.000000,15.000000) -- (9.000000,0.000000);
\begin{scope}
\draw[fill=white] (9.000000, 0.000000) circle(3.000000pt);
\clip (9.000000, 0.000000) circle(3.000000pt);
\draw (6.000000, 0.000000) -- (12.000000, 0.000000);
\draw (9.000000, -3.000000) -- (9.000000, 3.000000);
\end{scope}
\filldraw (9.000000, 15.000000) circle(1.500000pt);
\draw (27.000000,30.000000) -- (27.000000,15.000000);
\begin{scope}
\draw[fill=white] (27.000000, 30.000000) circle(3.000000pt);
\clip (27.000000, 30.000000) circle(3.000000pt);
\draw (24.000000, 30.000000) -- (30.000000, 30.000000);
\draw (27.000000, 27.000000) -- (27.000000, 33.000000);
\end{scope}
\filldraw (27.000000, 15.000000) circle(1.500000pt);
\draw (59.500000,45.000000) -- (59.500000,30.000000);
\begin{scope}
\draw[fill=white] (59.500000, 37.500000) +(-45.000000:24.748737pt and 19.091883pt) -- +(45.000000:24.748737pt and 19.091883pt) -- +(135.000000:24.748737pt and 19.091883pt) -- +(225.000000:24.748737pt and 19.091883pt) -- cycle;
\clip (59.500000, 37.500000) +(-45.000000:24.748737pt and 19.091883pt) -- +(45.000000:24.748737pt and 19.091883pt) -- +(135.000000:24.748737pt and 19.091883pt) -- +(225.000000:24.748737pt and 19.091883pt) -- cycle;
\draw (59.500000, 37.500000) node {$R_{ZZ}(\frac{\pi}{4})$};
\end{scope}
\begin{scope}[color=blue]
\begin{scope}[color=blue]
\begin{scope}
\draw[fill=white] (95.000000, 45.000000) +(-45.000000:8.485281pt and 8.485281pt) -- +(45.000000:8.485281pt and 8.485281pt) -- +(135.000000:8.485281pt and 8.485281pt) -- +(225.000000:8.485281pt and 8.485281pt) -- cycle;
\clip (95.000000, 45.000000) +(-45.000000:8.485281pt and 8.485281pt) -- +(45.000000:8.485281pt and 8.485281pt) -- +(135.000000:8.485281pt and 8.485281pt) -- +(225.000000:8.485281pt and 8.485281pt) -- cycle;
\draw (95.000000, 45.000000) node {$Z$};
\end{scope}
\end{scope}
\end{scope}
\begin{scope}[color=blue]
\begin{scope}[color=blue]
\begin{scope}
\draw[fill=white] (95.000000, 30.000000) +(-45.000000:8.485281pt and 8.485281pt) -- +(45.000000:8.485281pt and 8.485281pt) -- +(135.000000:8.485281pt and 8.485281pt) -- +(225.000000:8.485281pt and 8.485281pt) -- cycle;
\clip (95.000000, 30.000000) +(-45.000000:8.485281pt and 8.485281pt) -- +(45.000000:8.485281pt and 8.485281pt) -- +(135.000000:8.485281pt and 8.485281pt) -- +(225.000000:8.485281pt and 8.485281pt) -- cycle;
\draw (95.000000, 30.000000) node {$Z$};
\end{scope}
\end{scope}
\end{scope}
\draw (133.000000,45.000000) -- (133.000000,15.000000);
\begin{scope}
\draw[fill=white] (133.000000, 37.500000) +(-45.000000:28.284271pt and 19.091883pt) -- +(45.000000:28.284271pt and 19.091883pt) -- +(135.000000:28.284271pt and 19.091883pt) -- +(225.000000:28.284271pt and 19.091883pt) -- cycle;
\clip (133.000000, 37.500000) +(-45.000000:28.284271pt and 19.091883pt) -- +(45.000000:28.284271pt and 19.091883pt) -- +(135.000000:28.284271pt and 19.091883pt) -- +(225.000000:28.284271pt and 19.091883pt) -- cycle;
\draw (133.000000, 37.500000) node {$R_{ZZ}(\frac{-\pi}{2})$};
\end{scope}
\filldraw (133.000000, 15.000000) circle(1.500000pt);
\begin{scope}[color=red]
\begin{scope}[color=red]
\begin{scope}
\draw[fill=white] (171.000000, 45.000000) +(-45.000000:8.485281pt and 8.485281pt) -- +(45.000000:8.485281pt and 8.485281pt) -- +(135.000000:8.485281pt and 8.485281pt) -- +(225.000000:8.485281pt and 8.485281pt) -- cycle;
\clip (171.000000, 45.000000) +(-45.000000:8.485281pt and 8.485281pt) -- +(45.000000:8.485281pt and 8.485281pt) -- +(135.000000:8.485281pt and 8.485281pt) -- +(225.000000:8.485281pt and 8.485281pt) -- cycle;
\draw (171.000000, 45.000000) node {$Z$};
\end{scope}
\end{scope}
\end{scope}
\begin{scope}[color=red]
\begin{scope}[color=red]
\begin{scope}
\draw[fill=white] (171.000000, 30.000000) +(-45.000000:8.485281pt and 8.485281pt) -- +(45.000000:8.485281pt and 8.485281pt) -- +(135.000000:8.485281pt and 8.485281pt) -- +(225.000000:8.485281pt and 8.485281pt) -- cycle;
\clip (171.000000, 30.000000) +(-45.000000:8.485281pt and 8.485281pt) -- +(45.000000:8.485281pt and 8.485281pt) -- +(135.000000:8.485281pt and 8.485281pt) -- +(225.000000:8.485281pt and 8.485281pt) -- cycle;
\draw (171.000000, 30.000000) node {$Z$};
\end{scope}
\end{scope}
\end{scope}
\begin{scope}[color=red]
\begin{scope}[color=red]
\begin{scope}
\draw[fill=white] (171.000000, 15.000000) +(-45.000000:8.485281pt and 8.485281pt) -- +(45.000000:8.485281pt and 8.485281pt) -- +(135.000000:8.485281pt and 8.485281pt) -- +(225.000000:8.485281pt and 8.485281pt) -- cycle;
\clip (171.000000, 15.000000) +(-45.000000:8.485281pt and 8.485281pt) -- +(45.000000:8.485281pt and 8.485281pt) -- +(135.000000:8.485281pt and 8.485281pt) -- +(225.000000:8.485281pt and 8.485281pt) -- cycle;
\draw (171.000000, 15.000000) node {$Z$};
\end{scope}
\end{scope}
\end{scope}
\draw (192.000000,30.000000) -- (192.000000,15.000000);
\begin{scope}
\draw[fill=white] (192.000000, 30.000000) circle(3.000000pt);
\clip (192.000000, 30.000000) circle(3.000000pt);
\draw (189.000000, 30.000000) -- (195.000000, 30.000000);
\draw (192.000000, 27.000000) -- (192.000000, 33.000000);
\end{scope}
\filldraw (192.000000, 15.000000) circle(1.500000pt);
\draw (192.000000,60.000000) -- (192.000000,45.000000);
\begin{scope}
\draw[fill=white] (192.000000, 60.000000) circle(3.000000pt);
\clip (192.000000, 60.000000) circle(3.000000pt);
\draw (189.000000, 60.000000) -- (195.000000, 60.000000);
\draw (192.000000, 57.000000) -- (192.000000, 63.000000);
\end{scope}
\filldraw (192.000000, 45.000000) circle(1.500000pt);
\draw (213.000000,15.000000) -- (213.000000,0.000000);
\begin{scope}
\draw[fill=white] (213.000000, 0.000000) circle(3.000000pt);
\clip (213.000000, 0.000000) circle(3.000000pt);
\draw (210.000000, 0.000000) -- (216.000000, 0.000000);
\draw (213.000000, -3.000000) -- (213.000000, 3.000000);
\end{scope}
\filldraw (213.000000, 15.000000) circle(1.500000pt);
\draw[fill=white] (207.000000, 56.000000) -- (215.000000,56.000000) arc (-90:90:4.000000pt) -- (207.000000,64.000000) -- cycle;
\draw (213.000000, 60.000000) node {{\scriptsize $Z$}};
\draw[fill=white] (231.000000, 11.000000) -- (239.000000,11.000000) arc (-90:90:4.000000pt) -- (231.000000,19.000000) -- cycle;
\draw (237.000000, 15.000000) node {{\scriptsize $X$}};
\draw[fill=white] (231.000000, -4.000000) -- (239.000000,-4.000000) arc (-90:90:4.000000pt) -- (231.000000,4.000000) -- cycle;
\draw (237.000000, 0.000000) node {{\scriptsize $Z$}};
\end{tikzpicture}
    \caption{Measurement of a gauge operator of $R_{ZZ_{q_b,q_i}}(\pi/4)$. The flag $a_0$ will detect the $ZZ$ error shown in blue, but the flag circuitry itself introduces an undetected error shown in red.}
    \label{fig:nonft_piby4}
\end{figure}

However, this circuit is not fault-tolerant because the ${C_{a_0}R_{ZZ}}_{q_i,q_b}(-\nicefrac{\pi}{2})$ gate can itself introduce a logical $ZZ$ error on $\I{k}$ (shown in red) with an additional $Z$ component on $a_0$ so that it is not detected by the measurement of that flag ancilla. Since this is a three-qubit gate, a priori, this error cannot be ruled out (and in fact, it will occur in our implementation of this gate). Consequently, we must measure a gauge operator which anticommutes with the spackle of this error. Since our construction hinges on measuring gauge operators formed by propagating $X$ operators through multi-controlled small angle rotations, we derive the gauge operator for a generic multi-controlled many-body $Z$ rotation:

\begin{equation}\label{eq:gauge multicontrolled}
    \begin{tikzpicture}[scale=1.000000,x=1pt,y=1pt]
\filldraw[color=white] (0.000000, -7.500000) rectangle (179.000000, 82.500000);
\draw[color=black] (0.000000,75.000000) -- (179.000000,75.000000);
\draw[color=black] (0.000000,75.000000) node[left] {$q_0$};
\draw[color=black] (0.000000,60.000000) node[anchor=mid east] {$\vdots$};
\draw[color=black] (0.000000,45.000000) -- (179.000000,45.000000);
\draw[color=black] (0.000000,45.000000) node[left] {$q_{r-1}$};
\draw[color=black] (0.000000,30.000000) -- (179.000000,30.000000);
\draw[color=black] (0.000000,30.000000) node[left] {$c_{0}$};
\draw[color=black] (0.000000,15.000000) node[anchor=mid east] {$\vdots$};
\draw[color=black] (0.000000,0.000000) -- (179.000000,0.000000);
\draw[color=black] (0.000000,0.000000) node[left] {$c_{m-1}$};
\draw (12.000000,75.000000) -- (12.000000,0.000000);
\begin{scope}
\draw[fill=white] (12.000000, 37.500000) +(-45.000000:8.485281pt and 61.518290pt) -- +(45.000000:8.485281pt and 61.518290pt) -- +(135.000000:8.485281pt and 61.518290pt) -- +(225.000000:8.485281pt and 61.518290pt) -- cycle;
\clip (12.000000, 37.500000) +(-45.000000:8.485281pt and 61.518290pt) -- +(45.000000:8.485281pt and 61.518290pt) -- +(135.000000:8.485281pt and 61.518290pt) -- +(225.000000:8.485281pt and 61.518290pt) -- cycle;
\draw (12.000000, 37.500000) node {$G$};
\end{scope}
\draw (50.000000,75.000000) -- (50.000000,0.000000);
\begin{scope}
\draw[fill=white] (50.000000, 60.000000) +(-45.000000:28.284271pt and 29.698485pt) -- +(45.000000:28.284271pt and 29.698485pt) -- +(135.000000:28.284271pt and 29.698485pt) -- +(225.000000:28.284271pt and 29.698485pt) -- cycle;
\clip (50.000000, 60.000000) +(-45.000000:28.284271pt and 29.698485pt) -- +(45.000000:28.284271pt and 29.698485pt) -- +(135.000000:28.284271pt and 29.698485pt) -- +(225.000000:28.284271pt and 29.698485pt) -- cycle;
\draw (50.000000, 60.000000) node {$R_{Z...Z}(\theta)$};
\end{scope}
\filldraw (50.000000, 30.000000) circle(1.500000pt);
\filldraw (50.000000, 0.000000) circle(1.500000pt);
\begin{scope}
\draw[fill=white] (88.000000, 75.000000) +(-45.000000:8.485281pt and 8.485281pt) -- +(45.000000:8.485281pt and 8.485281pt) -- +(135.000000:8.485281pt and 8.485281pt) -- +(225.000000:8.485281pt and 8.485281pt) -- cycle;
\clip (88.000000, 75.000000) +(-45.000000:8.485281pt and 8.485281pt) -- +(45.000000:8.485281pt and 8.485281pt) -- +(135.000000:8.485281pt and 8.485281pt) -- +(225.000000:8.485281pt and 8.485281pt) -- cycle;
\draw (88.000000, 75.000000) node {$X$};
\end{scope}
\draw[fill=white,color=white] (106.000000, -6.000000) rectangle (121.000000, 81.000000);
\draw (113.500000, 37.500000) node {$=$};
\draw (153.000000,75.000000) -- (153.000000,0.000000);
\begin{scope}
\draw[fill=white] (153.000000, 60.000000) +(-45.000000:28.284271pt and 29.698485pt) -- +(45.000000:28.284271pt and 29.698485pt) -- +(135.000000:28.284271pt and 29.698485pt) -- +(225.000000:28.284271pt and 29.698485pt) -- cycle;
\clip (153.000000, 60.000000) +(-45.000000:28.284271pt and 29.698485pt) -- +(45.000000:28.284271pt and 29.698485pt) -- +(135.000000:28.284271pt and 29.698485pt) -- +(225.000000:28.284271pt and 29.698485pt) -- cycle;
\draw (153.000000, 60.000000) node {$R_{Z...Z}(\theta)$};
\end{scope}
\filldraw (153.000000, 30.000000) circle(1.500000pt);
\filldraw (153.000000, 0.000000) circle(1.500000pt);
\draw[color=black] (179.000000,60.000000) node[anchor=mid west] {$\vdots$};
\draw[color=black] (179.000000,15.000000) node[anchor=mid west] {$\vdots$};
\end{tikzpicture}.
\end{equation}

implying that 

\begin{equation}
    \begin{tikzpicture}[scale=1.000000,x=1pt,y=1pt]
\filldraw[color=white] (0.000000, -7.500000) rectangle (332.000000, 82.500000);
\draw[color=black] (0.000000,75.000000) -- (332.000000,75.000000);
\draw[color=black] (0.000000,75.000000) node[left] {$q_0$};
\draw[color=black] (0.000000,60.000000) node[anchor=mid east] {$\vdots$};
\draw[color=black] (0.000000,45.000000) -- (332.000000,45.000000);
\draw[color=black] (0.000000,45.000000) node[left] {$q_{r-1}$};
\draw[color=black] (0.000000,30.000000) -- (332.000000,30.000000);
\draw[color=black] (0.000000,30.000000) node[left] {$c_{0}$};
\draw[color=black] (0.000000,15.000000) node[anchor=mid east] {$\vdots$};
\draw[color=black] (0.000000,0.000000) -- (332.000000,0.000000);
\draw[color=black] (0.000000,0.000000) node[left] {$c_{m-1}$};
\draw (12.000000,75.000000) -- (12.000000,0.000000);
\begin{scope}
\draw[fill=white] (12.000000, 37.500000) +(-45.000000:8.485281pt and 61.518290pt) -- +(45.000000:8.485281pt and 61.518290pt) -- +(135.000000:8.485281pt and 61.518290pt) -- +(225.000000:8.485281pt and 61.518290pt) -- cycle;
\clip (12.000000, 37.500000) +(-45.000000:8.485281pt and 61.518290pt) -- +(45.000000:8.485281pt and 61.518290pt) -- +(135.000000:8.485281pt and 61.518290pt) -- +(225.000000:8.485281pt and 61.518290pt) -- cycle;
\draw (12.000000, 37.500000) node {$G$};
\end{scope}
\draw[fill=white,color=white] (30.000000, -6.000000) rectangle (45.000000, 81.000000);
\draw (37.500000, 37.500000) node {$=$};
\draw (82.000000,75.000000) -- (82.000000,0.000000);
\begin{scope}
\draw[fill=white] (82.000000, 60.000000) +(-45.000000:35.355339pt and 29.698485pt) -- +(45.000000:35.355339pt and 29.698485pt) -- +(135.000000:35.355339pt and 29.698485pt) -- +(225.000000:35.355339pt and 29.698485pt) -- cycle;
\clip (82.000000, 60.000000) +(-45.000000:35.355339pt and 29.698485pt) -- +(45.000000:35.355339pt and 29.698485pt) -- +(135.000000:35.355339pt and 29.698485pt) -- +(225.000000:35.355339pt and 29.698485pt) -- cycle;
\draw (82.000000, 60.000000) node {$R_{Z...Z}(\pm \frac{\pi}{2^l})$};
\end{scope}
\filldraw (82.000000, 30.000000) circle(1.500000pt);
\filldraw (82.000000, 0.000000) circle(1.500000pt);
\begin{scope}
\draw[fill=white] (125.000000, 75.000000) +(-45.000000:8.485281pt and 8.485281pt) -- +(45.000000:8.485281pt and 8.485281pt) -- +(135.000000:8.485281pt and 8.485281pt) -- +(225.000000:8.485281pt and 8.485281pt) -- cycle;
\clip (125.000000, 75.000000) +(-45.000000:8.485281pt and 8.485281pt) -- +(45.000000:8.485281pt and 8.485281pt) -- +(135.000000:8.485281pt and 8.485281pt) -- +(225.000000:8.485281pt and 8.485281pt) -- cycle;
\draw (125.000000, 75.000000) node {$X$};
\end{scope}
\draw (170.500000,75.000000) -- (170.500000,0.000000);
\begin{scope}
\draw[fill=white] (170.500000, 60.000000) +(-45.000000:38.890873pt and 29.698485pt) -- +(45.000000:38.890873pt and 29.698485pt) -- +(135.000000:38.890873pt and 29.698485pt) -- +(225.000000:38.890873pt and 29.698485pt) -- cycle;
\clip (170.500000, 60.000000) +(-45.000000:38.890873pt and 29.698485pt) -- +(45.000000:38.890873pt and 29.698485pt) -- +(135.000000:38.890873pt and 29.698485pt) -- +(225.000000:38.890873pt and 29.698485pt) -- cycle;
\draw (170.500000, 60.000000) node {$R_{Z...Z}(\mp \frac{\pi}{2^l})$};
\end{scope}
\filldraw (170.500000, 30.000000) circle(1.500000pt);
\filldraw (170.500000, 0.000000) circle(1.500000pt);
\draw[fill=white,color=white] (210.000000, -6.000000) rectangle (225.000000, 81.000000);
\draw (217.500000, 37.500000) node {$=$};
\begin{scope}
\draw[fill=white] (243.000000, 75.000000) +(-45.000000:8.485281pt and 8.485281pt) -- +(45.000000:8.485281pt and 8.485281pt) -- +(135.000000:8.485281pt and 8.485281pt) -- +(225.000000:8.485281pt and 8.485281pt) -- cycle;
\clip (243.000000, 75.000000) +(-45.000000:8.485281pt and 8.485281pt) -- +(45.000000:8.485281pt and 8.485281pt) -- +(135.000000:8.485281pt and 8.485281pt) -- +(225.000000:8.485281pt and 8.485281pt) -- cycle;
\draw (243.000000, 75.000000) node {$X$};
\end{scope}
\draw (293.500000,75.000000) -- (293.500000,0.000000);
\begin{scope}
\draw[fill=white] (293.500000, 60.000000) +(-45.000000:45.961941pt and 29.698485pt) -- +(45.000000:45.961941pt and 29.698485pt) -- +(135.000000:45.961941pt and 29.698485pt) -- +(225.000000:45.961941pt and 29.698485pt) -- cycle;
\clip (293.500000, 60.000000) +(-45.000000:45.961941pt and 29.698485pt) -- +(45.000000:45.961941pt and 29.698485pt) -- +(135.000000:45.961941pt and 29.698485pt) -- +(225.000000:45.961941pt and 29.698485pt) -- cycle;
\draw (293.500000, 60.000000) node {$R_{Z...Z}(\mp \frac{\pi}{2^{l-1}})$};
\end{scope}
\filldraw (293.500000, 30.000000) circle(1.500000pt);
\filldraw (293.500000, 0.000000) circle(1.500000pt);
\draw[color=black] (332.000000,60.000000) node[anchor=mid west] {$\vdots$};
\draw[color=black] (332.000000,15.000000) node[anchor=mid west] {$\vdots$};
\end{tikzpicture}.
\end{equation}

\begin{figure}[!htbp]
\centering
    \begin{tikzpicture}[scale=1.000000,x=1pt,y=1pt]
\filldraw[color=white] (0.000000, -7.500000) rectangle (293.000000, 97.500000);
\draw[color=black] (0.000000,90.000000) -- (221.000000,90.000000);
\draw[color=black] (221.000000,89.500000) -- (293.000000,89.500000);
\draw[color=black] (221.000000,90.500000) -- (293.000000,90.500000);
\draw[color=black] (0.000000,90.000000) node[left] {$a_z\colon|0\rangle$};
\draw[color=black] (0.000000,75.000000) -- (293.000000,75.000000);
\draw[color=black] (0.000000,75.000000) node[left] {$q_b$};
\draw[color=black] (0.000000,60.000000) -- (293.000000,60.000000);
\draw[color=black] (0.000000,60.000000) node[left] {$q_i$};
\draw[color=black] (0.000000,45.000000) -- (281.000000,45.000000);
\draw[color=black] (281.000000,44.500000) -- (293.000000,44.500000);
\draw[color=black] (281.000000,45.500000) -- (293.000000,45.500000);
\draw[color=black] (0.000000,45.000000) node[left] {$a_0\colon|+\rangle$};
\draw[color=black] (0.000000,30.000000) -- (281.000000,30.000000);
\draw[color=black] (281.000000,29.500000) -- (293.000000,29.500000);
\draw[color=black] (281.000000,30.500000) -- (293.000000,30.500000);
\draw[color=black] (0.000000,30.000000) node[left] {$f_0\colon|0\rangle$};
\draw[color=black] (0.000000,15.000000) -- (281.000000,15.000000);
\draw[color=black] (281.000000,14.500000) -- (293.000000,14.500000);
\draw[color=black] (281.000000,15.500000) -- (293.000000,15.500000);
\draw[color=black] (0.000000,15.000000) node[left] {$a_1\colon|+\rangle$};
\draw[color=black] (0.000000,0.000000) -- (281.000000,0.000000);
\draw[color=black] (281.000000,-0.500000) -- (293.000000,-0.500000);
\draw[color=black] (281.000000,0.500000) -- (293.000000,0.500000);
\draw[color=black] (0.000000,0.000000) node[left] {$f_1\colon|0\rangle$};
\draw (9.000000,90.000000) -- (9.000000,75.000000);
\begin{scope}
\draw[fill=white] (9.000000, 90.000000) circle(3.000000pt);
\clip (9.000000, 90.000000) circle(3.000000pt);
\draw (6.000000, 90.000000) -- (12.000000, 90.000000);
\draw (9.000000, 87.000000) -- (9.000000, 93.000000);
\end{scope}
\filldraw (9.000000, 75.000000) circle(1.500000pt);
\draw (9.000000,45.000000) -- (9.000000,30.000000);
\begin{scope}
\draw[fill=white] (9.000000, 30.000000) circle(3.000000pt);
\clip (9.000000, 30.000000) circle(3.000000pt);
\draw (6.000000, 30.000000) -- (12.000000, 30.000000);
\draw (9.000000, 27.000000) -- (9.000000, 33.000000);
\end{scope}
\filldraw (9.000000, 45.000000) circle(1.500000pt);
\draw (9.000000,15.000000) -- (9.000000,0.000000);
\begin{scope}
\draw[fill=white] (9.000000, 0.000000) circle(3.000000pt);
\clip (9.000000, 0.000000) circle(3.000000pt);
\draw (6.000000, 0.000000) -- (12.000000, 0.000000);
\draw (9.000000, -3.000000) -- (9.000000, 3.000000);
\end{scope}
\filldraw (9.000000, 15.000000) circle(1.500000pt);
\draw (27.000000,60.000000) -- (27.000000,45.000000);
\begin{scope}
\draw[fill=white] (27.000000, 60.000000) circle(3.000000pt);
\clip (27.000000, 60.000000) circle(3.000000pt);
\draw (24.000000, 60.000000) -- (30.000000, 60.000000);
\draw (27.000000, 57.000000) -- (27.000000, 63.000000);
\end{scope}
\filldraw (27.000000, 45.000000) circle(1.500000pt);
\draw (59.500000,75.000000) -- (59.500000,60.000000);
\begin{scope}
\draw[fill=white] (59.500000, 67.500000) +(-45.000000:24.748737pt and 19.091883pt) -- +(45.000000:24.748737pt and 19.091883pt) -- +(135.000000:24.748737pt and 19.091883pt) -- +(225.000000:24.748737pt and 19.091883pt) -- cycle;
\clip (59.500000, 67.500000) +(-45.000000:24.748737pt and 19.091883pt) -- +(45.000000:24.748737pt and 19.091883pt) -- +(135.000000:24.748737pt and 19.091883pt) -- +(225.000000:24.748737pt and 19.091883pt) -- cycle;
\draw (59.500000, 67.500000) node {$R_{ZZ}(\frac{\pi}{4})$};
\end{scope}
\draw (92.000000,60.000000) -- (92.000000,15.000000);
\begin{scope}
\draw[fill=white] (92.000000, 60.000000) circle(3.000000pt);
\clip (92.000000, 60.000000) circle(3.000000pt);
\draw (89.000000, 60.000000) -- (95.000000, 60.000000);
\draw (92.000000, 57.000000) -- (92.000000, 63.000000);
\end{scope}
\filldraw (92.000000, 15.000000) circle(1.500000pt);
\draw (127.000000,75.000000) -- (127.000000,45.000000);
\begin{scope}
\draw[fill=white] (127.000000, 67.500000) +(-45.000000:28.284271pt and 19.091883pt) -- +(45.000000:28.284271pt and 19.091883pt) -- +(135.000000:28.284271pt and 19.091883pt) -- +(225.000000:28.284271pt and 19.091883pt) -- cycle;
\clip (127.000000, 67.500000) +(-45.000000:28.284271pt and 19.091883pt) -- +(45.000000:28.284271pt and 19.091883pt) -- +(135.000000:28.284271pt and 19.091883pt) -- +(225.000000:28.284271pt and 19.091883pt) -- cycle;
\draw (127.000000, 67.500000) node {$R_{ZZ}(\frac{-\pi}{2})$};
\end{scope}
\filldraw (127.000000, 45.000000) circle(1.500000pt);
\draw (165.000000,75.000000) -- (165.000000,15.000000);
\begin{scope}
\draw[fill=white] (165.000000, 75.000000) +(-45.000000:8.485281pt and 8.485281pt) -- +(45.000000:8.485281pt and 8.485281pt) -- +(135.000000:8.485281pt and 8.485281pt) -- +(225.000000:8.485281pt and 8.485281pt) -- cycle;
\clip (165.000000, 75.000000) +(-45.000000:8.485281pt and 8.485281pt) -- +(45.000000:8.485281pt and 8.485281pt) -- +(135.000000:8.485281pt and 8.485281pt) -- +(225.000000:8.485281pt and 8.485281pt) -- cycle;
\draw (165.000000, 75.000000) node {$Z$};
\end{scope}
\filldraw (165.000000, 45.000000) circle(1.500000pt);
\filldraw (165.000000, 15.000000) circle(1.500000pt);
\draw (193.000000,60.000000) -- (193.000000,15.000000);
\begin{scope}
\draw[fill=white] (193.000000, 60.000000) +(-45.000000:14.142136pt and 8.485281pt) -- +(45.000000:14.142136pt and 8.485281pt) -- +(135.000000:14.142136pt and 8.485281pt) -- +(225.000000:14.142136pt and 8.485281pt) -- cycle;
\clip (193.000000, 60.000000) +(-45.000000:14.142136pt and 8.485281pt) -- +(45.000000:14.142136pt and 8.485281pt) -- +(135.000000:14.142136pt and 8.485281pt) -- +(225.000000:14.142136pt and 8.485281pt) -- cycle;
\draw (193.000000, 60.000000) node {$-iZ$};
\end{scope}
\filldraw (193.000000, 45.000000) circle(1.500000pt);
\filldraw (193.000000, 15.000000) circle(1.500000pt);
\draw (193.000000,90.000000) -- (193.000000,75.000000);
\begin{scope}
\draw[fill=white] (193.000000, 90.000000) circle(3.000000pt);
\clip (193.000000, 90.000000) circle(3.000000pt);
\draw (190.000000, 90.000000) -- (196.000000, 90.000000);
\draw (193.000000, 87.000000) -- (193.000000, 93.000000);
\end{scope}
\filldraw (193.000000, 75.000000) circle(1.500000pt);
\draw (221.000000,60.000000) -- (221.000000,15.000000);
\begin{scope}
\draw[fill=white] (221.000000, 60.000000) circle(3.000000pt);
\clip (221.000000, 60.000000) circle(3.000000pt);
\draw (218.000000, 60.000000) -- (224.000000, 60.000000);
\draw (221.000000, 57.000000) -- (221.000000, 63.000000);
\end{scope}
\filldraw (221.000000, 15.000000) circle(1.500000pt);
\draw[fill=white] (215.000000, 86.000000) -- (223.000000,86.000000) arc (-90:90:4.000000pt) -- (215.000000,94.000000) -- cycle;
\draw (221.000000, 90.000000) node {{\scriptsize $Z$}};
\draw (242.000000,60.000000) -- (242.000000,45.000000);
\begin{scope}
\draw[fill=white] (242.000000, 60.000000) circle(3.000000pt);
\clip (242.000000, 60.000000) circle(3.000000pt);
\draw (239.000000, 60.000000) -- (245.000000, 60.000000);
\draw (242.000000, 57.000000) -- (242.000000, 63.000000);
\end{scope}
\filldraw (242.000000, 45.000000) circle(1.500000pt);
\draw (260.000000,45.000000) -- (260.000000,30.000000);
\begin{scope}
\draw[fill=white] (260.000000, 30.000000) circle(3.000000pt);
\clip (260.000000, 30.000000) circle(3.000000pt);
\draw (257.000000, 30.000000) -- (263.000000, 30.000000);
\draw (260.000000, 27.000000) -- (260.000000, 33.000000);
\end{scope}
\filldraw (260.000000, 45.000000) circle(1.500000pt);
\draw (260.000000,15.000000) -- (260.000000,0.000000);
\begin{scope}
\draw[fill=white] (260.000000, 0.000000) circle(3.000000pt);
\clip (260.000000, 0.000000) circle(3.000000pt);
\draw (257.000000, 0.000000) -- (263.000000, 0.000000);
\draw (260.000000, -3.000000) -- (260.000000, 3.000000);
\end{scope}
\filldraw (260.000000, 15.000000) circle(1.500000pt);
\draw[fill=white] (275.000000, 41.000000) -- (283.000000,41.000000) arc (-90:90:4.000000pt) -- (275.000000,49.000000) -- cycle;
\draw (281.000000, 45.000000) node {{\scriptsize $X$}};
\draw[fill=white] (275.000000, 26.000000) -- (283.000000,26.000000) arc (-90:90:4.000000pt) -- (275.000000,34.000000) -- cycle;
\draw (281.000000, 30.000000) node {{\scriptsize $Z$}};
\draw[fill=white] (275.000000, 11.000000) -- (283.000000,11.000000) arc (-90:90:4.000000pt) -- (275.000000,19.000000) -- cycle;
\draw (281.000000, 15.000000) node {{\scriptsize $X$}};
\draw[fill=white] (275.000000, -4.000000) -- (283.000000,-4.000000) arc (-90:90:4.000000pt) -- (275.000000,4.000000) -- cycle;
\draw (281.000000, 0.000000) node {{\scriptsize $Z$}};
\end{tikzpicture}
\caption{The fault-tolerant $T$ gate, $\overline{R_{ZZ_{q_i,q_b}}(\nicefrac{\pi}{4})}^{FT}$ produced by our construction in the iceberg code.}\label{fig: ft pi4}
\end{figure}
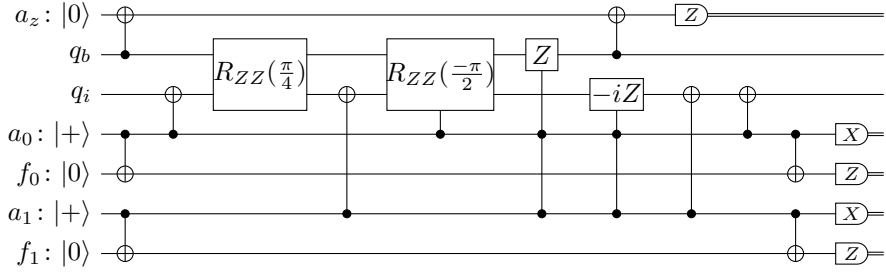

Therefore, we can detect $ZZ$ errors on ${C_{a_0}R_{ZZ}}_{q_i,q_b}(-\nicefrac{\pi}{2})$ by measuring the gauge operator consisting of an $X_{q_i}$ before the gate, followed by a $X_{q_i}{C_{a_0}R_{ZZ}}_{q_i,q_b}(\pi) = X_{q_i}C_{a_0}(-iZ_{q_i} Z_{q_b}) = X_{q_i}C_{a_0}(-iZ_{q_i}) C_{a_0}(Z_{q_b})$ after the gate. Replacing ${C_{a_0}R_{ZZ}}_{q_i,q_b}(-\nicefrac{\pi}{2})$ with this flagged version in the $\nicefrac{\pi}{4}$ circuit, we now have the fault-tolerant $\overline{R_{ZZ}(\nicefrac{\pi}{4})_{q_i,q_b}}^{FT}$ circuit shown in Fig.~\ref{fig: ft pi4}.

Note that no further flags need to be added to the circuit because the $X$ operator on qubit $q_i$ propagates to an $R_{ZZ_{q_i,q_b}}(\pi) = -iZ_{q_i}Z_{q_b}$ operator, which can be factored into the tensor product of two single-qubit operators, which prevents a $ZZ$ error occurring on the data qubits from a single gate failure. This is the same reason that only one flag is required in the fault-tolerant $R_{ZZ}(\nicefrac{\pi}{2})$ circuit. In fact, any multi-controlled $R_{ZZ}(\pm\nicefrac{\pi}{2})$ acting on qubits $q_i$ and $q_b$ can be made fault-tolerant in the same way by one measuring one extra gauge operator derived from the propagation of an $X_{q_i}$ through this circuit element (only the number of controls in the multi-controlled $Z$ gates will vary).

To generalize the construction for the $\nicefrac{\pi}{4}$ rotation to smaller angle $\nicefrac{\pi}{2^l}$ rotations, we observe that in the $\nicefrac{\pi}{4}$ case we measured gauge operators of controlled $R_{ZZ}(\nicefrac{\pi}{2^l})$ rotations to detect $ZZ$ errors on the data qubits by further controlled-power-of-2 rotations until the angle of the rotation was $\pi$, in which case a single gate error could no longer induce an undetectable $ZZ$ error. We now formalize this in Definition~\ref{def:iceberg construction}

\begin{definition}\label{def:iceberg construction}
    Given a ${{C_{a_{m-1}, ... ,a_{0}}R_{ZZ}}_{q_i,q_b}(\pm\nicefrac{\pi}{2^l})}$ rotation, we denote by $\overline{{C_{a_{m-1}, ... ,a_{0}}R_{ZZ}}_{q_i,q_b}(\pm\nicefrac{\pi}{2^l})}^{FT, ZZ}$ a version of the circuit in which gauge operators are measured to detect $ZZ$ errors on the data qubits resulting from any single error within this gadget. $\overline{{C_{a_{m-1}, ... ,a_{0}}R_{ZZ}}_{q_i,q_b}(\pm\nicefrac{\pi}{2^l})}^{FT, ZZ}$ can be recursively defined by the equations
    \begin{equation}
        \begin{tikzpicture}[scale=1.000000,x=1pt,y=1pt]
\filldraw[color=white] (0.000000, -7.500000) rectangle (387.000000, 97.500000);
\draw[color=black] (0.000000,90.000000) -- (387.000000,90.000000);
\draw[color=black] (0.000000,90.000000) node[left] {$q_b$};
\draw[color=black] (0.000000,75.000000) -- (387.000000,75.000000);
\draw[color=black] (0.000000,75.000000) node[left] {$q_i$};
\draw[color=black] (0.000000,60.000000) -- (387.000000,60.000000);
\draw[color=black] (0.000000,60.000000) node[left] {$a_{0}$};
\draw[color=black] (0.000000,45.000000) node[anchor=mid east] {$\vdots$};
\draw[color=black] (0.000000,30.000000) -- (387.000000,30.000000);
\draw[color=black] (0.000000,30.000000) node[left] {$a_{m-1}$};
\draw[color=black] (123.000000,15.000000) -- (375.000000,15.000000);
\draw[color=black] (375.000000,14.500000) -- (387.000000,14.500000);
\draw[color=black] (375.000000,15.500000) -- (387.000000,15.500000);
\draw[color=black] (123.000000,0.000000) -- (375.000000,0.000000);
\draw[color=black] (375.000000,-0.500000) -- (387.000000,-0.500000);
\draw[color=black] (375.000000,0.500000) -- (387.000000,0.500000);
\draw (43.500000,90.000000) -- (43.500000,30.000000);
\begin{scope}
\draw[fill=white] (43.500000, 82.500000) +(-45.000000:53.033009pt and 19.091883pt) -- +(45.000000:53.033009pt and 19.091883pt) -- +(135.000000:53.033009pt and 19.091883pt) -- +(225.000000:53.033009pt and 19.091883pt) -- cycle;
\clip (43.500000, 82.500000) +(-45.000000:53.033009pt and 19.091883pt) -- +(45.000000:53.033009pt and 19.091883pt) -- +(135.000000:53.033009pt and 19.091883pt) -- +(225.000000:53.033009pt and 19.091883pt) -- cycle;
\draw (43.500000, 82.500000) node {$\overline{R_{ZZ}(\pm \frac{\pi}{2^l})}^{FT,ZZ}$};
\end{scope}
\filldraw (43.500000, 60.000000) circle(1.500000pt);
\filldraw (43.500000, 30.000000) circle(1.500000pt);
\draw[fill=white,color=white] (93.000000, -6.000000) rectangle (108.000000, 96.000000);
\draw (100.500000, 45.000000) node {$=$};
\filldraw[color=white] (120.000000, 12.000000) rectangle (126.000000, 18.000000);
\draw (126.000000, 12.000000) -- (126.000000, 18.000000);
\draw (123.000000, 15.000000) node {$\scriptstyle{\hspace{-30 pt}a_{m}\colon|+\rangle}$};
\filldraw[color=white] (120.000000, -3.000000) rectangle (126.000000, 3.000000);
\draw (126.000000, -3.000000) -- (126.000000, 3.000000);
\draw (123.000000, 0.000000) node {$\scriptstyle{\hspace{-30 pt}f_{m}\colon|0\rangle}$};
\draw (141.000000,15.000000) -- (141.000000,0.000000);
\begin{scope}
\draw[fill=white] (141.000000, 0.000000) circle(3.000000pt);
\clip (141.000000, 0.000000) circle(3.000000pt);
\draw (138.000000, 0.000000) -- (144.000000, 0.000000);
\draw (141.000000, -3.000000) -- (141.000000, 3.000000);
\end{scope}
\filldraw (141.000000, 15.000000) circle(1.500000pt);
\draw (159.000000,75.000000) -- (159.000000,15.000000);
\begin{scope}
\draw[fill=white] (159.000000, 75.000000) circle(3.000000pt);
\clip (159.000000, 75.000000) circle(3.000000pt);
\draw (156.000000, 75.000000) -- (162.000000, 75.000000);
\draw (159.000000, 72.000000) -- (159.000000, 78.000000);
\end{scope}
\filldraw (159.000000, 15.000000) circle(1.500000pt);
\draw (199.000000,90.000000) -- (199.000000,30.000000);
\begin{scope}
\draw[fill=white] (199.000000, 82.500000) +(-45.000000:35.355339pt and 19.091883pt) -- +(45.000000:35.355339pt and 19.091883pt) -- +(135.000000:35.355339pt and 19.091883pt) -- +(225.000000:35.355339pt and 19.091883pt) -- cycle;
\clip (199.000000, 82.500000) +(-45.000000:35.355339pt and 19.091883pt) -- +(45.000000:35.355339pt and 19.091883pt) -- +(135.000000:35.355339pt and 19.091883pt) -- +(225.000000:35.355339pt and 19.091883pt) -- cycle;
\draw (199.000000, 82.500000) node {${R_{ZZ}(\pm \frac{\pi}{2^l})}$};
\end{scope}
\filldraw (199.000000, 60.000000) circle(1.500000pt);
\filldraw (199.000000, 30.000000) circle(1.500000pt);
\draw (278.500000,90.000000) -- (278.500000,15.000000);
\begin{scope}
\draw[fill=white] (278.500000, 82.500000) +(-45.000000:60.104076pt and 19.091883pt) -- +(45.000000:60.104076pt and 19.091883pt) -- +(135.000000:60.104076pt and 19.091883pt) -- +(225.000000:60.104076pt and 19.091883pt) -- cycle;
\clip (278.500000, 82.500000) +(-45.000000:60.104076pt and 19.091883pt) -- +(45.000000:60.104076pt and 19.091883pt) -- +(135.000000:60.104076pt and 19.091883pt) -- +(225.000000:60.104076pt and 19.091883pt) -- cycle;
\draw (278.500000, 82.500000) node {$\overline{R_{ZZ}(\mp \frac{\pi}{2^l-1})}^{FT,ZZ}$};
\end{scope}
\filldraw (278.500000, 60.000000) circle(1.500000pt);
\filldraw (278.500000, 30.000000) circle(1.500000pt);
\filldraw (278.500000, 15.000000) circle(1.500000pt);
\draw (336.000000,75.000000) -- (336.000000,15.000000);
\begin{scope}
\draw[fill=white] (336.000000, 75.000000) circle(3.000000pt);
\clip (336.000000, 75.000000) circle(3.000000pt);
\draw (333.000000, 75.000000) -- (339.000000, 75.000000);
\draw (336.000000, 72.000000) -- (336.000000, 78.000000);
\end{scope}
\filldraw (336.000000, 15.000000) circle(1.500000pt);
\draw (354.000000,15.000000) -- (354.000000,0.000000);
\begin{scope}
\draw[fill=white] (354.000000, 0.000000) circle(3.000000pt);
\clip (354.000000, 0.000000) circle(3.000000pt);
\draw (351.000000, 0.000000) -- (357.000000, 0.000000);
\draw (354.000000, -3.000000) -- (354.000000, 3.000000);
\end{scope}
\filldraw (354.000000, 15.000000) circle(1.500000pt);
\draw[fill=white] (369.000000, 11.000000) -- (377.000000,11.000000) arc (-90:90:4.000000pt) -- (369.000000,19.000000) -- cycle;
\draw (375.000000, 15.000000) node {{\scriptsize $X$}};
\draw[fill=white] (369.000000, -4.000000) -- (377.000000,-4.000000) arc (-90:90:4.000000pt) -- (369.000000,4.000000) -- cycle;
\draw (375.000000, 0.000000) node {{\scriptsize $Z$}};
\draw[color=black] (387.000000,45.000000) node[anchor=mid west] {$\vdots$};
\end{tikzpicture} 
    \end{equation}
    for $l > 1$, the recursive step, and
    \begin{equation}
        \centering
        \begin{tikzpicture}[scale=1.000000,x=1pt,y=1pt]
\filldraw[color=white] (0.000000, -7.500000) rectangle (356.000000, 97.500000);
\draw[color=black] (0.000000,90.000000) -- (356.000000,90.000000);
\draw[color=black] (0.000000,90.000000) node[left] {$q_b$};
\draw[color=black] (0.000000,75.000000) -- (356.000000,75.000000);
\draw[color=black] (0.000000,75.000000) node[left] {$q_i$};
\draw[color=black] (0.000000,60.000000) -- (356.000000,60.000000);
\draw[color=black] (0.000000,60.000000) node[left] {$a_{0}$};
\draw[color=black] (0.000000,45.000000) node[anchor=mid east] {$\vdots$};
\draw[color=black] (0.000000,30.000000) -- (356.000000,30.000000);
\draw[color=black] (0.000000,30.000000) node[left] {$a_{m-1}$};
\draw[color=black] (123.000000,15.000000) -- (344.000000,15.000000);
\draw[color=black] (344.000000,14.500000) -- (356.000000,14.500000);
\draw[color=black] (344.000000,15.500000) -- (356.000000,15.500000);
\draw[color=black] (123.000000,0.000000) -- (344.000000,0.000000);
\draw[color=black] (344.000000,-0.500000) -- (356.000000,-0.500000);
\draw[color=black] (344.000000,0.500000) -- (356.000000,0.500000);
\draw (43.500000,90.000000) -- (43.500000,30.000000);
\begin{scope}
\draw[fill=white] (43.500000, 82.500000) +(-45.000000:53.033009pt and 19.091883pt) -- +(45.000000:53.033009pt and 19.091883pt) -- +(135.000000:53.033009pt and 19.091883pt) -- +(225.000000:53.033009pt and 19.091883pt) -- cycle;
\clip (43.500000, 82.500000) +(-45.000000:53.033009pt and 19.091883pt) -- +(45.000000:53.033009pt and 19.091883pt) -- +(135.000000:53.033009pt and 19.091883pt) -- +(225.000000:53.033009pt and 19.091883pt) -- cycle;
\draw (43.500000, 82.500000) node {$\overline{R_{ZZ}(\pm \frac{\pi}{2})}^{FT,ZZ}$};
\end{scope}
\filldraw (43.500000, 60.000000) circle(1.500000pt);
\filldraw (43.500000, 30.000000) circle(1.500000pt);
\draw[fill=white,color=white] (93.000000, -6.000000) rectangle (108.000000, 96.000000);
\draw (100.500000, 45.000000) node {$=$};
\filldraw[color=white] (120.000000, 12.000000) rectangle (126.000000, 18.000000);
\draw (126.000000, 12.000000) -- (126.000000, 18.000000);
\draw (123.000000, 15.000000) node {$\scriptstyle{\hspace{-30 pt}a_{m}\colon|+\rangle}$};
\filldraw[color=white] (120.000000, -3.000000) rectangle (126.000000, 3.000000);
\draw (126.000000, -3.000000) -- (126.000000, 3.000000);
\draw (123.000000, 0.000000) node {$\scriptstyle{\hspace{-30 pt}f_{m}\colon|0\rangle}$};
\draw (141.000000,15.000000) -- (141.000000,0.000000);
\begin{scope}
\draw[fill=white] (141.000000, 0.000000) circle(3.000000pt);
\clip (141.000000, 0.000000) circle(3.000000pt);
\draw (138.000000, 0.000000) -- (144.000000, 0.000000);
\draw (141.000000, -3.000000) -- (141.000000, 3.000000);
\end{scope}
\filldraw (141.000000, 15.000000) circle(1.500000pt);
\draw (159.000000,75.000000) -- (159.000000,15.000000);
\begin{scope}
\draw[fill=white] (159.000000, 75.000000) circle(3.000000pt);
\clip (159.000000, 75.000000) circle(3.000000pt);
\draw (156.000000, 75.000000) -- (162.000000, 75.000000);
\draw (159.000000, 72.000000) -- (159.000000, 78.000000);
\end{scope}
\filldraw (159.000000, 15.000000) circle(1.500000pt);
\draw (199.000000,90.000000) -- (199.000000,30.000000);
\begin{scope}
\draw[fill=white] (199.000000, 82.500000) +(-45.000000:35.355339pt and 19.091883pt) -- +(45.000000:35.355339pt and 19.091883pt) -- +(135.000000:35.355339pt and 19.091883pt) -- +(225.000000:35.355339pt and 19.091883pt) -- cycle;
\clip (199.000000, 82.500000) +(-45.000000:35.355339pt and 19.091883pt) -- +(45.000000:35.355339pt and 19.091883pt) -- +(135.000000:35.355339pt and 19.091883pt) -- +(225.000000:35.355339pt and 19.091883pt) -- cycle;
\draw (199.000000, 82.500000) node {${R_{ZZ}(\pm \frac{\pi}{2})}$};
\end{scope}
\filldraw (199.000000, 60.000000) circle(1.500000pt);
\filldraw (199.000000, 30.000000) circle(1.500000pt);
\draw (242.000000,90.000000) -- (242.000000,15.000000);
\begin{scope}
\draw[fill=white] (242.000000, 90.000000) +(-45.000000:8.485281pt and 8.485281pt) -- +(45.000000:8.485281pt and 8.485281pt) -- +(135.000000:8.485281pt and 8.485281pt) -- +(225.000000:8.485281pt and 8.485281pt) -- cycle;
\clip (242.000000, 90.000000) +(-45.000000:8.485281pt and 8.485281pt) -- +(45.000000:8.485281pt and 8.485281pt) -- +(135.000000:8.485281pt and 8.485281pt) -- +(225.000000:8.485281pt and 8.485281pt) -- cycle;
\draw (242.000000, 90.000000) node {$Z$};
\end{scope}
\filldraw (242.000000, 60.000000) circle(1.500000pt);
\filldraw (242.000000, 30.000000) circle(1.500000pt);
\filldraw (242.000000, 15.000000) circle(1.500000pt);
\draw (275.000000,75.000000) -- (275.000000,15.000000);
\begin{scope}
\draw[fill=white] (275.000000, 75.000000) +(-45.000000:21.213203pt and 8.485281pt) -- +(45.000000:21.213203pt and 8.485281pt) -- +(135.000000:21.213203pt and 8.485281pt) -- +(225.000000:21.213203pt and 8.485281pt) -- cycle;
\clip (275.000000, 75.000000) +(-45.000000:21.213203pt and 8.485281pt) -- +(45.000000:21.213203pt and 8.485281pt) -- +(135.000000:21.213203pt and 8.485281pt) -- +(225.000000:21.213203pt and 8.485281pt) -- cycle;
\draw (275.000000, 75.000000) node {$\pm iZ$};
\end{scope}
\filldraw (275.000000, 60.000000) circle(1.500000pt);
\filldraw (275.000000, 30.000000) circle(1.500000pt);
\filldraw (275.000000, 15.000000) circle(1.500000pt);
\draw (305.000000,75.000000) -- (305.000000,15.000000);
\begin{scope}
\draw[fill=white] (305.000000, 75.000000) circle(3.000000pt);
\clip (305.000000, 75.000000) circle(3.000000pt);
\draw (302.000000, 75.000000) -- (308.000000, 75.000000);
\draw (305.000000, 72.000000) -- (305.000000, 78.000000);
\end{scope}
\filldraw (305.000000, 15.000000) circle(1.500000pt);
\draw (323.000000,15.000000) -- (323.000000,0.000000);
\begin{scope}
\draw[fill=white] (323.000000, 0.000000) circle(3.000000pt);
\clip (323.000000, 0.000000) circle(3.000000pt);
\draw (320.000000, 0.000000) -- (326.000000, 0.000000);
\draw (323.000000, -3.000000) -- (323.000000, 3.000000);
\end{scope}
\filldraw (323.000000, 15.000000) circle(1.500000pt);
\draw[fill=white] (338.000000, 11.000000) -- (346.000000,11.000000) arc (-90:90:4.000000pt) -- (338.000000,19.000000) -- cycle;
\draw (344.000000, 15.000000) node {{\scriptsize $X$}};
\draw[fill=white] (338.000000, -4.000000) -- (346.000000,-4.000000) arc (-90:90:4.000000pt) -- (338.000000,4.000000) -- cycle;
\draw (344.000000, 0.000000) node {{\scriptsize $Z$}};
\draw[color=black] (356.000000,45.000000) node[anchor=mid west] {$\vdots$};
\end{tikzpicture}
    \end{equation}
    for $l=1,$ the base case.
\end{definition}

We observe that the above construction is fault-tolerant to $ZZ$ errors in the sense that there will be some detector for the circuit (if we include the iceberg stabilizers as well) which will be deterministically flipped if the resulting error configuration from any circuit failure has a $ZZ$ component on the data qubits. An $X$ error on an ancilla $a_i$ occurring midway through the circuit will deterministically flip the flag $f_i$, and an $X_q$ error on a data qubit $q$ occurring before or during the circuit will propagate to some diagonal operator times $X_q$, which will deterministically flip the sign of the $Z$ stabilizer of $\I{k}.$ $ZZ$ errors on the data qubits occurring during a multi-controlled $ZZ$ rotation will be caught by the flag circuit for that rotation, except for the $\overline{{C_{a_{m-1}, ... ,a_{0}}R_{ZZ}}_{q_i,q_b}(\pm\nicefrac{\pi}{2})}^{FT, ZZ}$, which cannot induce a $ZZ$ error on the data qubits from a single circuit-failure.

Lastly, we add an additional gauge operator to check for $XX$ or $YY$ errors:
\begin{definition}
    We denote our circuit for a fault-tolerant $\nicefrac{\pi}{2^l}$ rotation around logical qubit $i$ by $\overline{R_Z(\nicefrac{\pi}{2^l})}^{log, FT}_i := \overline{{R_{ZZ}}_{q_i,q_b}(\pm\nicefrac{\pi}{2^{l}})}^{FT}$. We have that 
    \begin{equation}
        \begin{tikzpicture}[scale=1.000000,x=1pt,y=1pt]
\filldraw[color=white] (0.000000, -7.500000) rectangle (274.000000, 37.500000);
\draw[color=black] (118.000000,30.000000) -- (262.000000,30.000000);
\draw[color=black] (262.000000,29.500000) -- (274.000000,29.500000);
\draw[color=black] (262.000000,30.500000) -- (274.000000,30.500000);
\draw[color=black] (0.000000,15.000000) -- (274.000000,15.000000);
\draw[color=black] (0.000000,15.000000) node[left] {$q_b$};
\draw[color=black] (0.000000,0.000000) -- (274.000000,0.000000);
\draw[color=black] (0.000000,0.000000) node[left] {$q_i$};
\draw (41.000000,15.000000) -- (41.000000,0.000000);
\begin{scope}
\draw[fill=white] (41.000000, 7.500000) +(-45.000000:49.497475pt and 19.091883pt) -- +(45.000000:49.497475pt and 19.091883pt) -- +(135.000000:49.497475pt and 19.091883pt) -- +(225.000000:49.497475pt and 19.091883pt) -- cycle;
\clip (41.000000, 7.500000) +(-45.000000:49.497475pt and 19.091883pt) -- +(45.000000:49.497475pt and 19.091883pt) -- +(135.000000:49.497475pt and 19.091883pt) -- +(225.000000:49.497475pt and 19.091883pt) -- cycle;
\draw (41.000000, 7.500000) node {$\overline{R_{ZZ}(\pm \frac{\pi}{2^l})}^{FT}$};
\end{scope}
\draw[fill=white,color=white] (88.000000, -6.000000) rectangle (103.000000, 36.000000);
\draw (95.500000, 15.000000) node {$=$};
\filldraw[color=white] (115.000000, 27.000000) rectangle (121.000000, 33.000000);
\draw (121.000000, 27.000000) -- (121.000000, 33.000000);
\draw (118.000000, 30.000000) node {$\scriptstyle{\hspace{-30 pt}a_{z}\colon|0\rangle}$};
\draw (136.000000,30.000000) -- (136.000000,15.000000);
\begin{scope}
\draw[fill=white] (136.000000, 30.000000) circle(3.000000pt);
\clip (136.000000, 30.000000) circle(3.000000pt);
\draw (133.000000, 30.000000) -- (139.000000, 30.000000);
\draw (136.000000, 27.000000) -- (136.000000, 33.000000);
\end{scope}
\filldraw (136.000000, 15.000000) circle(1.500000pt);
\draw (188.500000,15.000000) -- (188.500000,0.000000);
\begin{scope}
\draw[fill=white] (188.500000, 7.500000) +(-45.000000:53.033009pt and 19.091883pt) -- +(45.000000:53.033009pt and 19.091883pt) -- +(135.000000:53.033009pt and 19.091883pt) -- +(225.000000:53.033009pt and 19.091883pt) -- cycle;
\clip (188.500000, 7.500000) +(-45.000000:53.033009pt and 19.091883pt) -- +(45.000000:53.033009pt and 19.091883pt) -- +(135.000000:53.033009pt and 19.091883pt) -- +(225.000000:53.033009pt and 19.091883pt) -- cycle;
\draw (188.500000, 7.500000) node {$\overline{R_{ZZ}(\pm \frac{\pi}{2^l})}^{FT,ZZ}$};
\end{scope}
\draw (241.000000,30.000000) -- (241.000000,15.000000);
\begin{scope}
\draw[fill=white] (241.000000, 30.000000) circle(3.000000pt);
\clip (241.000000, 30.000000) circle(3.000000pt);
\draw (238.000000, 30.000000) -- (244.000000, 30.000000);
\draw (241.000000, 27.000000) -- (241.000000, 33.000000);
\end{scope}
\filldraw (241.000000, 15.000000) circle(1.500000pt);
\draw[fill=white] (256.000000, 26.000000) -- (264.000000,26.000000) arc (-90:90:4.000000pt) -- (256.000000,34.000000) -- cycle;
\draw (262.000000, 30.000000) node {{\scriptsize $Z$}};
\end{tikzpicture}
    \end{equation}
\end{definition}

\begin{figure}[!htpb]
    \centering
    \includegraphics[width=0.5\linewidth]{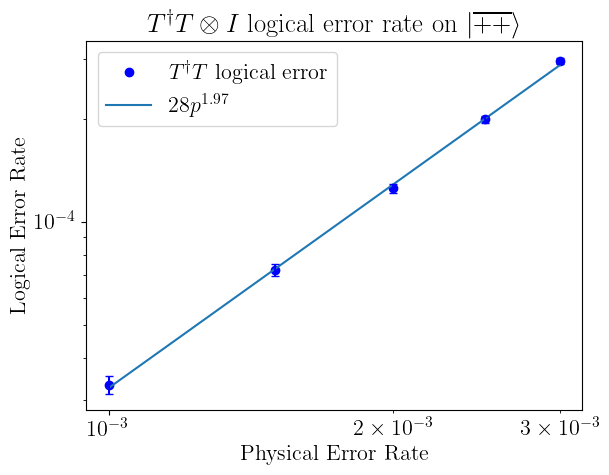}\includegraphics[width=0.5\linewidth]{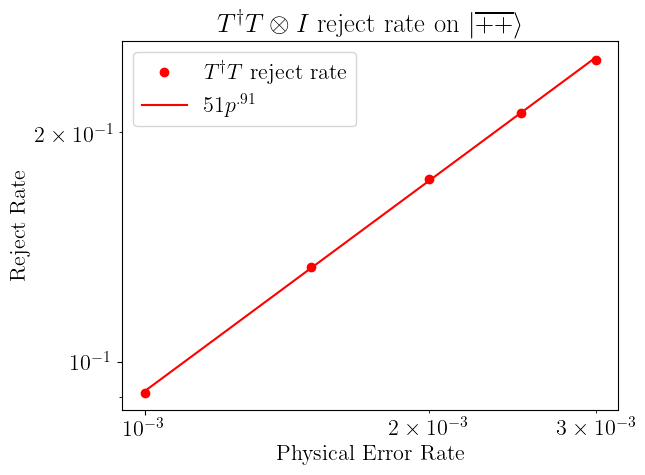}
    \caption{State-vector simulations varying the noise strength parameter $p$ for the fault-tolerant $T$ gate on the first logical qubit of a $\nkd{4}{2}{2}$ code applied to the $\ket{\overline{++}}$ input state. To decrease the sample complexity required to estimate the logical error rate, we benchmark the noisy rotation of the first qubit by $\nicefrac{\pi}{4}$ followed by its inverse, returning to $\ket{\overline{++}}$. After performing these two gates, we extract Z syndromes using flagged syndrome extraction and X syndromes in a final destructive measurement. We post-select on any non-trivial syndromes or flags. Circuits were written in Guppy~\cite{koch2025} and state-vector simulations were performed in Selene~\cite{Selene}.}
    \label{fig:ft pi4 iceberg sim}
\end{figure}

\begin{figure}
    \centering
    \includegraphics[width=0.5\linewidth]{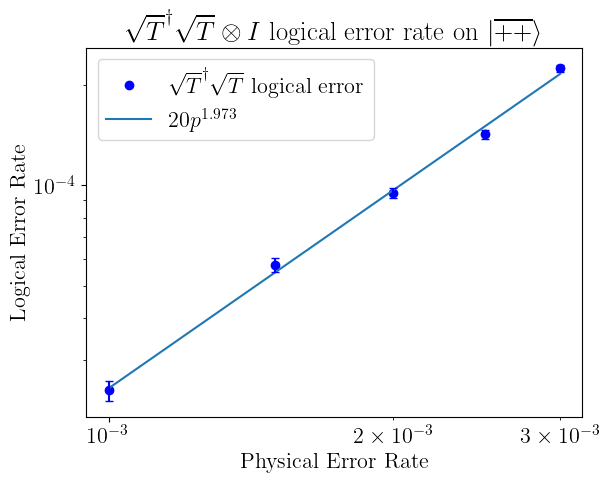}\includegraphics[width=0.5\linewidth]{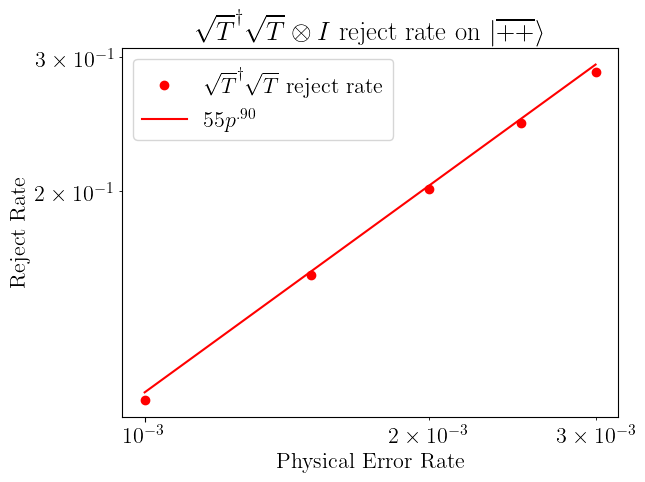}
    \caption{State-vector simulations varying the noise strength parameter $p$ for the fault-tolerant $\sqrt{T}$ gate on the first logical qubit of a $\nkd{4}{2}{2}$ code applied to the $\ket{\overline{++}}$ input state. We apply the inverse rotation as above.}
    \label{fig:ft pi8 iceberg sim}
\end{figure}

Simulations are given for the $\overline{R_{\overline{Z}}(\nicefrac{\pi}{4})}^{log,FT}_1 $ and $\overline{R_{\overline{Z}}(\nicefrac{\pi}{8})}^{log,FT}_1$ circuits in Figs.~\ref{fig:ft pi4 iceberg sim} and~\ref{fig:ft pi8 iceberg sim} demonstrating the correct scaling with the strength of the noise parameter $p$. In all simulations, our error model will consist of the application of two-qubit depolarizing noise with strength $p$ after every two-qubit gate, as well as measurement error with probability $p$ applied to all measurements.

Now we calculate the size of the circuit for $\overline{R_{\overline{Z}}(\nicefrac{\pi}{2^l})}^{log,FT}_i$. In this circuit, there is a $j$-controlled gate for every $j \leq l$ (and two for $j = l$). The cost of our circuit mainly comes from our decomposition of these $j$-controlled gates. If one uses the single-ancilla-qubit construction of these multi-controlled gates with depth $O(\log j)$ and gate count $O(j)$ \cite{Kitaev2002}, then the overall gate count of the circuit will be $O(\sum_{j=1}^{l}j) = O(l^2)$ and the depth will be $O(\sum_{j=1}^l \log j) = O(l \log l).$ However, we observe that the controlled gates in our circuits have a simple structure that we can exploit. For $j < l$, the $j$-controlled gate is followed by a $j+1$-controlled gate sharing all the controls of the $j$-controlled gate, having one additional control. Suppose that we had a garbage qubit $g_{j-2}$ that stored the value of the $j$ controls just before the $j+1$-controlled gate with an additional control $c_{j+1}$. Then we could perform a Toffoli controlled on $g_{j-2}$, $c_{j+1}$, targeting an additional garbage qubit $g_{j-1}$ initialized to $\ket{0}$. We can then apply our next controlled rotation with $g_{j-1}$ as the control. After all the controlled rotations are performed and before the $CX$ gates, we can undo the ladder of Toffolis on the garbage qubits. Unlike in constructions where the garbage qubits can be in an arbitrary state, we will require that they are initialized to $\ket{0}$ so that we can measure them in the Z basis after the ladder is completed to check for $X$ hook errors in the ladder. The depth and gate count of this ladder is $O(l).$ The garbage qubits and toffoli ladder are shown explicitly in the circuit for $\overline{R_{\overline{Z}}(\nicefrac{\pi}{8})}^{log,FT}_i$ in Fig.~\ref{fig: ft pi8}

Finally, we note that essentially the same circuits apply to rotations of the form $x_0.x_{1}x_{2}\ldots x_{l}$ with the same overheads in $l.$ The recursion is modified so that a controlled $x_0.x_{1}x_{2}\ldots x_{l}$ rotation is gauged via a controlled $x_1.x_{2}x_{3}\ldots x_{l}$ rotation with one less digit of precision. The base case corresponds to a controlled $x_0.x_{1}$ rotation. We give the precise circuits in Definition~\ref{def:iceberg decimal construction} below.

\begin{definition}\label{def:iceberg decimal construction}
    Given a ${{C_{a_{m-1}, ... ,a_{0}}R_{ZZ}}_{q_i,q_b}(\pm \pi(x_0.x_{1}\ldots x_{l}))}$ rotation, we denote by \\ $\overline{{C_{a_{m-1}, ... ,a_{0}}R_{ZZ}}_{q_i,q_b}(\pm \pi(x_0.x_{1}\ldots x_{l}))}^{FT, ZZ}$ a version of the circuit in which gauge operators are measured to detect $ZZ$ errors on the data qubits resulting from any single error within this gadget as before. \\$\overline{{C_{a_{m-1}, ... ,a_{0}}R_{ZZ}}_{q_i,q_b}(\pm \pi(x_0.x_{1}\ldots x_{l}))}^{FT, ZZ}$ can be recursively defined by the equations
    \begin{equation}
        \hspace{-1.1cm} \resizebox{\textwidth}{!}{\begin{tikzpicture}[scale=1.000000,x=1pt,y=1pt]
\filldraw[color=white] (0.000000, -7.500000) rectangle (542.000000, 97.500000);
\draw[color=black] (0.000000,90.000000) -- (542.000000,90.000000);
\draw[color=black] (0.000000,90.000000) node[left] {$q_b$};
\draw[color=black] (0.000000,75.000000) -- (542.000000,75.000000);
\draw[color=black] (0.000000,75.000000) node[left] {$q_i$};
\draw[color=black] (0.000000,60.000000) -- (542.000000,60.000000);
\draw[color=black] (0.000000,60.000000) node[left] {$a_{0}$};
\draw[color=black] (0.000000,45.000000) node[anchor=mid east] {$\vdots$};
\draw[color=black] (0.000000,30.000000) -- (542.000000,30.000000);
\draw[color=black] (0.000000,30.000000) node[left] {$a_{m-1}$};
\draw[color=black] (173.000000,15.000000) -- (530.000000,15.000000);
\draw[color=black] (530.000000,14.500000) -- (542.000000,14.500000);
\draw[color=black] (530.000000,15.500000) -- (542.000000,15.500000);
\draw[color=black] (173.000000,0.000000) -- (530.000000,0.000000);
\draw[color=black] (530.000000,-0.500000) -- (542.000000,-0.500000);
\draw[color=black] (530.000000,0.500000) -- (542.000000,0.500000);
\draw (68.500000,90.000000) -- (68.500000,30.000000);
\begin{scope}
\draw[fill=white] (68.500000, 82.500000) +(-45.000000:88.388348pt and 19.091883pt) -- +(45.000000:88.388348pt and 19.091883pt) -- +(135.000000:88.388348pt and 19.091883pt) -- +(225.000000:88.388348pt and 19.091883pt) -- cycle;
\clip (68.500000, 82.500000) +(-45.000000:88.388348pt and 19.091883pt) -- +(45.000000:88.388348pt and 19.091883pt) -- +(135.000000:88.388348pt and 19.091883pt) -- +(225.000000:88.388348pt and 19.091883pt) -- cycle;
\draw (68.500000, 82.500000) node {$\overline{R_{ZZ}(\pm \pi(x_0.x_{1}\ldots x_{l}))}^{FT,ZZ}$};
\end{scope}
\filldraw (68.500000, 60.000000) circle(1.500000pt);
\filldraw (68.500000, 30.000000) circle(1.500000pt);
\draw[fill=white,color=white] (143.000000, -6.000000) rectangle (158.000000, 96.000000);
\draw (150.500000, 45.000000) node {$=$};
\filldraw[color=white] (170.000000, 12.000000) rectangle (176.000000, 18.000000);
\draw (176.000000, 12.000000) -- (176.000000, 18.000000);
\draw (173.000000, 15.000000) node {$\scriptstyle{\hspace{-30 pt}a_{m}\colon|+\rangle}$};
\filldraw[color=white] (170.000000, -3.000000) rectangle (176.000000, 3.000000);
\draw (176.000000, -3.000000) -- (176.000000, 3.000000);
\draw (173.000000, 0.000000) node {$\scriptstyle{\hspace{-30 pt}f_{m}\colon|0\rangle}$};
\draw (191.000000,15.000000) -- (191.000000,0.000000);
\begin{scope}
\draw[fill=white] (191.000000, 0.000000) circle(3.000000pt);
\clip (191.000000, 0.000000) circle(3.000000pt);
\draw (188.000000, 0.000000) -- (194.000000, 0.000000);
\draw (191.000000, -3.000000) -- (191.000000, 3.000000);
\end{scope}
\filldraw (191.000000, 15.000000) circle(1.500000pt);
\draw (209.000000,75.000000) -- (209.000000,15.000000);
\begin{scope}
\draw[fill=white] (209.000000, 75.000000) circle(3.000000pt);
\clip (209.000000, 75.000000) circle(3.000000pt);
\draw (206.000000, 75.000000) -- (212.000000, 75.000000);
\draw (209.000000, 72.000000) -- (209.000000, 78.000000);
\end{scope}
\filldraw (209.000000, 15.000000) circle(1.500000pt);
\draw (279.000000,90.000000) -- (279.000000,30.000000);
\begin{scope}
\draw[fill=white] (279.000000, 82.500000) +(-45.000000:77.781746pt and 19.091883pt) -- +(45.000000:77.781746pt and 19.091883pt) -- +(135.000000:77.781746pt and 19.091883pt) -- +(225.000000:77.781746pt and 19.091883pt) -- cycle;
\clip (279.000000, 82.500000) +(-45.000000:77.781746pt and 19.091883pt) -- +(45.000000:77.781746pt and 19.091883pt) -- +(135.000000:77.781746pt and 19.091883pt) -- +(225.000000:77.781746pt and 19.091883pt) -- cycle;
\draw (279.000000, 82.500000) node {${R_{ZZ}(\pm \pi(x_0.x_{1}\ldots x_{l}))}$};
\end{scope}
\filldraw (279.000000, 60.000000) circle(1.500000pt);
\filldraw (279.000000, 30.000000) circle(1.500000pt);
\draw (411.000000,90.000000) -- (411.000000,15.000000);
\begin{scope}
\draw[fill=white] (411.000000, 82.500000) +(-45.000000:91.923882pt and 19.091883pt) -- +(45.000000:91.923882pt and 19.091883pt) -- +(135.000000:91.923882pt and 19.091883pt) -- +(225.000000:91.923882pt and 19.091883pt) -- cycle;
\clip (411.000000, 82.500000) +(-45.000000:91.923882pt and 19.091883pt) -- +(45.000000:91.923882pt and 19.091883pt) -- +(135.000000:91.923882pt and 19.091883pt) -- +(225.000000:91.923882pt and 19.091883pt) -- cycle;
\draw (411.000000, 82.500000) node {$\overline{R_{ZZ}(\mp \pi(x_1.x_2\ldots x_{l}))}^{FT,ZZ}$};
\end{scope}
\filldraw (411.000000, 60.000000) circle(1.500000pt);
\filldraw (411.000000, 30.000000) circle(1.500000pt);
\filldraw (411.000000, 15.000000) circle(1.500000pt);
\draw (491.000000,75.000000) -- (491.000000,15.000000);
\begin{scope}
\draw[fill=white] (491.000000, 75.000000) circle(3.000000pt);
\clip (491.000000, 75.000000) circle(3.000000pt);
\draw (488.000000, 75.000000) -- (494.000000, 75.000000);
\draw (491.000000, 72.000000) -- (491.000000, 78.000000);
\end{scope}
\filldraw (491.000000, 15.000000) circle(1.500000pt);
\draw (509.000000,15.000000) -- (509.000000,0.000000);
\begin{scope}
\draw[fill=white] (509.000000, 0.000000) circle(3.000000pt);
\clip (509.000000, 0.000000) circle(3.000000pt);
\draw (506.000000, 0.000000) -- (512.000000, 0.000000);
\draw (509.000000, -3.000000) -- (509.000000, 3.000000);
\end{scope}
\filldraw (509.000000, 15.000000) circle(1.500000pt);
\draw[fill=white] (524.000000, 11.000000) -- (532.000000,11.000000) arc (-90:90:4.000000pt) -- (524.000000,19.000000) -- cycle;
\draw (530.000000, 15.000000) node {{\scriptsize $X$}};
\draw[fill=white] (524.000000, -4.000000) -- (532.000000,-4.000000) arc (-90:90:4.000000pt) -- (524.000000,4.000000) -- cycle;
\draw (530.000000, 0.000000) node {{\scriptsize $Z$}};
\draw[color=black] (542.000000,45.000000) node[anchor=mid west] {$\vdots$};
\end{tikzpicture}}
    \end{equation}
    for $l > 1$, the recursive step, and
    \begin{equation}
        \centering
        \begin{tikzpicture}[scale=1.000000,x=1pt,y=1pt]
\filldraw[color=white] (0.000000, -7.500000) rectangle (441.000000, 97.500000);
\draw[color=black] (0.000000,90.000000) -- (441.000000,90.000000);
\draw[color=black] (0.000000,90.000000) node[left] {$q_b$};
\draw[color=black] (0.000000,75.000000) -- (441.000000,75.000000);
\draw[color=black] (0.000000,75.000000) node[left] {$q_i$};
\draw[color=black] (0.000000,60.000000) -- (441.000000,60.000000);
\draw[color=black] (0.000000,60.000000) node[left] {$a_{0}$};
\draw[color=black] (0.000000,45.000000) node[anchor=mid east] {$\vdots$};
\draw[color=black] (0.000000,30.000000) -- (441.000000,30.000000);
\draw[color=black] (0.000000,30.000000) node[left] {$a_{m-1}$};
\draw[color=black] (153.000000,15.000000) -- (429.000000,15.000000);
\draw[color=black] (429.000000,14.500000) -- (441.000000,14.500000);
\draw[color=black] (429.000000,15.500000) -- (441.000000,15.500000);
\draw[color=black] (153.000000,0.000000) -- (429.000000,0.000000);
\draw[color=black] (429.000000,-0.500000) -- (441.000000,-0.500000);
\draw[color=black] (429.000000,0.500000) -- (441.000000,0.500000);
\draw (58.500000,90.000000) -- (58.500000,30.000000);
\begin{scope}
\draw[fill=white] (58.500000, 82.500000) +(-45.000000:74.246212pt and 19.091883pt) -- +(45.000000:74.246212pt and 19.091883pt) -- +(135.000000:74.246212pt and 19.091883pt) -- +(225.000000:74.246212pt and 19.091883pt) -- cycle;
\clip (58.500000, 82.500000) +(-45.000000:74.246212pt and 19.091883pt) -- +(45.000000:74.246212pt and 19.091883pt) -- +(135.000000:74.246212pt and 19.091883pt) -- +(225.000000:74.246212pt and 19.091883pt) -- cycle;
\draw (58.500000, 82.500000) node {$\overline{R_{ZZ}(\pm \pi(x_0.x_1))}^{FT,ZZ}$};
\end{scope}
\filldraw (58.500000, 60.000000) circle(1.500000pt);
\filldraw (58.500000, 30.000000) circle(1.500000pt);
\draw[fill=white,color=white] (123.000000, -6.000000) rectangle (138.000000, 96.000000);
\draw (130.500000, 45.000000) node {$=$};
\filldraw[color=white] (150.000000, 12.000000) rectangle (156.000000, 18.000000);
\draw (156.000000, 12.000000) -- (156.000000, 18.000000);
\draw (153.000000, 15.000000) node {$\scriptstyle{\hspace{-30 pt}a_{m}\colon|+\rangle}$};
\filldraw[color=white] (150.000000, -3.000000) rectangle (156.000000, 3.000000);
\draw (156.000000, -3.000000) -- (156.000000, 3.000000);
\draw (153.000000, 0.000000) node {$\scriptstyle{\hspace{-30 pt}f_{m}\colon|0\rangle}$};
\draw (171.000000,15.000000) -- (171.000000,0.000000);
\begin{scope}
\draw[fill=white] (171.000000, 0.000000) circle(3.000000pt);
\clip (171.000000, 0.000000) circle(3.000000pt);
\draw (168.000000, 0.000000) -- (174.000000, 0.000000);
\draw (171.000000, -3.000000) -- (171.000000, 3.000000);
\end{scope}
\filldraw (171.000000, 15.000000) circle(1.500000pt);
\draw (189.000000,75.000000) -- (189.000000,15.000000);
\begin{scope}
\draw[fill=white] (189.000000, 75.000000) circle(3.000000pt);
\clip (189.000000, 75.000000) circle(3.000000pt);
\draw (186.000000, 75.000000) -- (192.000000, 75.000000);
\draw (189.000000, 72.000000) -- (189.000000, 78.000000);
\end{scope}
\filldraw (189.000000, 15.000000) circle(1.500000pt);
\draw (241.500000,90.000000) -- (241.500000,30.000000);
\begin{scope}
\draw[fill=white] (241.500000, 82.500000) +(-45.000000:53.033009pt and 19.091883pt) -- +(45.000000:53.033009pt and 19.091883pt) -- +(135.000000:53.033009pt and 19.091883pt) -- +(225.000000:53.033009pt and 19.091883pt) -- cycle;
\clip (241.500000, 82.500000) +(-45.000000:53.033009pt and 19.091883pt) -- +(45.000000:53.033009pt and 19.091883pt) -- +(135.000000:53.033009pt and 19.091883pt) -- +(225.000000:53.033009pt and 19.091883pt) -- cycle;
\draw (241.500000, 82.500000) node {${R_{ZZ}(\pm \pi(x_0.x_1))}$};
\end{scope}
\filldraw (241.500000, 60.000000) circle(1.500000pt);
\filldraw (241.500000, 30.000000) circle(1.500000pt);
\draw (297.000000,90.000000) -- (297.000000,15.000000);
\begin{scope}
\draw[fill=white] (297.000000, 90.000000) +(-45.000000:8.485281pt and 8.485281pt) -- +(45.000000:8.485281pt and 8.485281pt) -- +(135.000000:8.485281pt and 8.485281pt) -- +(225.000000:8.485281pt and 8.485281pt) -- cycle;
\clip (297.000000, 90.000000) +(-45.000000:8.485281pt and 8.485281pt) -- +(45.000000:8.485281pt and 8.485281pt) -- +(135.000000:8.485281pt and 8.485281pt) -- +(225.000000:8.485281pt and 8.485281pt) -- cycle;
\draw (297.000000, 90.000000) node {$Z$};
\end{scope}
\filldraw (297.000000, 60.000000) circle(1.500000pt);
\filldraw (297.000000, 30.000000) circle(1.500000pt);
\filldraw (297.000000, 15.000000) circle(1.500000pt);
\draw (345.000000,75.000000) -- (345.000000,15.000000);
\begin{scope}
\draw[fill=white] (345.000000, 75.000000) +(-45.000000:42.426407pt and 8.485281pt) -- +(45.000000:42.426407pt and 8.485281pt) -- +(135.000000:42.426407pt and 8.485281pt) -- +(225.000000:42.426407pt and 8.485281pt) -- cycle;
\clip (345.000000, 75.000000) +(-45.000000:42.426407pt and 8.485281pt) -- +(45.000000:42.426407pt and 8.485281pt) -- +(135.000000:42.426407pt and 8.485281pt) -- +(225.000000:42.426407pt and 8.485281pt) -- cycle;
\draw (345.000000, 75.000000) node {$(-1^{x_0})(\pm iZ)$};
\end{scope}
\filldraw (345.000000, 60.000000) circle(1.500000pt);
\filldraw (345.000000, 30.000000) circle(1.500000pt);
\filldraw (345.000000, 15.000000) circle(1.500000pt);
\draw (390.000000,75.000000) -- (390.000000,15.000000);
\begin{scope}
\draw[fill=white] (390.000000, 75.000000) circle(3.000000pt);
\clip (390.000000, 75.000000) circle(3.000000pt);
\draw (387.000000, 75.000000) -- (393.000000, 75.000000);
\draw (390.000000, 72.000000) -- (390.000000, 78.000000);
\end{scope}
\filldraw (390.000000, 15.000000) circle(1.500000pt);
\draw (408.000000,15.000000) -- (408.000000,0.000000);
\begin{scope}
\draw[fill=white] (408.000000, 0.000000) circle(3.000000pt);
\clip (408.000000, 0.000000) circle(3.000000pt);
\draw (405.000000, 0.000000) -- (411.000000, 0.000000);
\draw (408.000000, -3.000000) -- (408.000000, 3.000000);
\end{scope}
\filldraw (408.000000, 15.000000) circle(1.500000pt);
\draw[fill=white] (423.000000, 11.000000) -- (431.000000,11.000000) arc (-90:90:4.000000pt) -- (423.000000,19.000000) -- cycle;
\draw (429.000000, 15.000000) node {{\scriptsize $X$}};
\draw[fill=white] (423.000000, -4.000000) -- (431.000000,-4.000000) arc (-90:90:4.000000pt) -- (423.000000,4.000000) -- cycle;
\draw (429.000000, 0.000000) node {{\scriptsize $Z$}};
\draw[color=black] (441.000000,45.000000) node[anchor=mid west] {$\vdots$};
\end{tikzpicture}
    \end{equation}
    for $l=1,$ the base case.
\end{definition}

\begin{figure}[h]
\centering
    \input{Figures/ZZ_pi_by_8.tikz}
\caption{The fault-tolerant $\sqrt{T}$ gate, $\overline{R_{\overline{Z}}(\nicefrac{\pi}{8})}^{log,FT}_1$ produced by our construction in the iceberg code.}\label{fig: ft pi8}
\end{figure}

\section{Generalization to other CSS codes}

\begin{figure}
    \centering
    \begin{tikzpicture}[scale=1.000000,x=1pt,y=1pt]
\filldraw[color=white] (0.000000, -7.500000) rectangle (150.000000, 97.500000);
\draw[color=black] (0.000000,90.000000) -- (150.000000,90.000000);
\draw[color=black] (0.000000,90.000000) node[left] {$q_0$};
\draw[color=black] (0.000000,75.000000) -- (150.000000,75.000000);
\draw[color=black] (0.000000,75.000000) node[left] {$q_1$};
\draw[color=black] (0.000000,60.000000) -- (150.000000,60.000000);
\draw[color=black] (0.000000,60.000000) node[left] {$q_2$};
\draw[color=black] (0.000000,45.000000) node[anchor=mid east] {$\vdots$};
\draw[color=black] (0.000000,30.000000) -- (150.000000,30.000000);
\draw[color=black] (0.000000,30.000000) node[left] {$q_{d-2}$};
\draw[color=black] (0.000000,15.000000) -- (150.000000,15.000000);
\draw[color=black] (0.000000,15.000000) node[left] {$q_{d-1}$};
\draw[color=black] (0.000000,0.000000) -- (150.000000,0.000000);
\draw[color=black] (0.000000,0.000000) node[left] {$a_{m-1}$};
\draw (9.000000,30.000000) -- (9.000000,15.000000);
\begin{scope}
\draw[fill=white] (9.000000, 30.000000) circle(3.000000pt);
\clip (9.000000, 30.000000) circle(3.000000pt);
\draw (6.000000, 30.000000) -- (12.000000, 30.000000);
\draw (9.000000, 27.000000) -- (9.000000, 33.000000);
\end{scope}
\filldraw (9.000000, 15.000000) circle(1.500000pt);
\draw (27.000000,75.000000) -- (27.000000,60.000000);
\begin{scope}
\draw[fill=white] (27.000000, 75.000000) circle(3.000000pt);
\clip (27.000000, 75.000000) circle(3.000000pt);
\draw (24.000000, 75.000000) -- (30.000000, 75.000000);
\draw (27.000000, 72.000000) -- (27.000000, 78.000000);
\end{scope}
\filldraw (27.000000, 60.000000) circle(1.500000pt);
\draw (45.000000,90.000000) -- (45.000000,75.000000);
\begin{scope}
\draw[fill=white] (45.000000, 90.000000) circle(3.000000pt);
\clip (45.000000, 90.000000) circle(3.000000pt);
\draw (42.000000, 90.000000) -- (48.000000, 90.000000);
\draw (45.000000, 87.000000) -- (45.000000, 93.000000);
\end{scope}
\filldraw (45.000000, 75.000000) circle(1.500000pt);
\begin{scope}
\draw[fill=white] (75.000000, 90.000000) +(-45.000000:21.213203pt and 8.485281pt) -- +(45.000000:21.213203pt and 8.485281pt) -- +(135.000000:21.213203pt and 8.485281pt) -- +(225.000000:21.213203pt and 8.485281pt) -- cycle;
\clip (75.000000, 90.000000) +(-45.000000:21.213203pt and 8.485281pt) -- +(45.000000:21.213203pt and 8.485281pt) -- +(135.000000:21.213203pt and 8.485281pt) -- +(225.000000:21.213203pt and 8.485281pt) -- cycle;
\draw (75.000000, 90.000000) node {${R_{Z}(\frac{\pi}{2^l})}$};
\end{scope}
\draw (105.000000,90.000000) -- (105.000000,75.000000);
\begin{scope}
\draw[fill=white] (105.000000, 90.000000) circle(3.000000pt);
\clip (105.000000, 90.000000) circle(3.000000pt);
\draw (102.000000, 90.000000) -- (108.000000, 90.000000);
\draw (105.000000, 87.000000) -- (105.000000, 93.000000);
\end{scope}
\filldraw (105.000000, 75.000000) circle(1.500000pt);
\draw (123.000000,75.000000) -- (123.000000,60.000000);
\begin{scope}
\draw[fill=white] (123.000000, 75.000000) circle(3.000000pt);
\clip (123.000000, 75.000000) circle(3.000000pt);
\draw (120.000000, 75.000000) -- (126.000000, 75.000000);
\draw (123.000000, 72.000000) -- (123.000000, 78.000000);
\end{scope}
\filldraw (123.000000, 60.000000) circle(1.500000pt);
\draw (141.000000,30.000000) -- (141.000000,15.000000);
\begin{scope}
\draw[fill=white] (141.000000, 30.000000) circle(3.000000pt);
\clip (141.000000, 30.000000) circle(3.000000pt);
\draw (138.000000, 30.000000) -- (144.000000, 30.000000);
\draw (141.000000, 27.000000) -- (141.000000, 33.000000);
\end{scope}
\filldraw (141.000000, 15.000000) circle(1.500000pt);
\draw[color=black] (150.000000,45.000000) node[anchor=mid west] {$\vdots$};
\end{tikzpicture}
    \caption{Non-fault-tolerant implementation of an $R_{\overline{Z}}(\frac{\pi}{2^l})$ on a distance-$d$ CSS code with logical Z operator $\bar{Z}_0 = Z_0 Z_1 ... Z_{d-1}$. }
    \label{fig:nonft arb}
\end{figure}
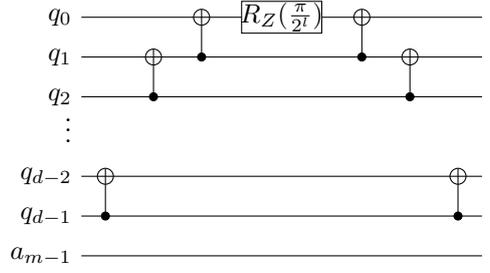

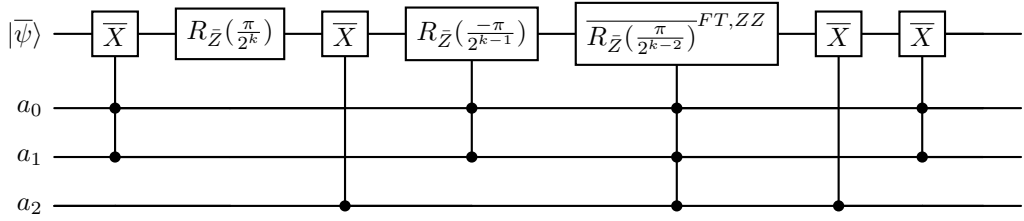
\begin{figure}[!htpb]
    \centering
    \begin{quantikz}
    \lstick{$\ket{\overline{\psi}}$} &\gate{\overline{X}}&\gate{R_{\bar{Z}}({\frac{\pi}{2^k}})}&\gate{\overline{X}}&\gate{R_{\bar{Z}}({\frac{-\pi}{2^{k-1}}})}&\gate{\overline{{R_{\bar{Z}}({\frac{\pi}{2^{k-2}}})}}^{FT,ZZ}}&\gate{\overline{X}}&\gate{\overline{X}}&\qw& \\
    \lstick{$a_0$} &\ctrl{-1}&\qw&\qw&\ctrl{-1}&\ctrl{0}&\qw&\ctrl{-1}&\qw& \\
    \lstick{$a_1$} &\ctrl{-1}&\qw&\qw&\ctrl{-1}&\ctrl{0}&\qw&\ctrl{-1}&\qw& \\
    \lstick{$a_2$} &\qw&\qw&\ctrl{-3}&\qw&\ctrl{-3}&\ctrl{-3}&\qw&\qw& \\
    \end{quantikz}
    \caption{Sequence of gauge operators to detect logical $Z$ errors in the general case. First, a logical $X$ operator is stored in an ancilla qubit $a_0$. Then, the non-fault-tolerant $R_{\overline{Z}}(-\nicefrac{\pi}{2^l})$ is performed, after which, an $R_{\overline{Z}}(-\nicefrac{\pi}{2^{l-1}})$ is stored into $a_0$. To detect logical errors induced by this $R_{\overline{Z}}(-\nicefrac{\pi}{2^{l-1}})$, we use an ancilla qubit $a_1$ which stores a $\bar{X}$ operator before the $R_{\overline{Z}}(-\nicefrac{\pi}{2^{l-1}})$ rotation followed by a logical $R_{\overline{Z}}(\frac{\pi}{2^{l-2}})$ rotation which we must make fault-tolerant, and so on. We denote the rest of this sequence by $\overline{{R_{\bar{Z}}({\frac{\pi}{2^{l-2}}})}}^{FT,ZZ}$ as before.}
    \label{fig:general ft gauge sequence}
\end{figure}

We now demonstrate that this recursive sequence of flagging circuits can also be applied to detect logical errors induced by non-fault-tolerant $R_{\overline{Z}}(\frac{\pi}{2^l})$ operators on other CSS codes with fault distance two. Suppose that $\mathcal{C}$ is a distance-$d$ CSS code, and relabel the qubits and logical operator so that the first logical $Z$ operator is given by $\bar{Z}_0 = Z_0 Z_1 ...Z_{d-1}.$ Then we could use the non-fault-tolerant implementation given in Fig.~\ref{fig:nonft arb} in which a single-qubit $R_{\overline{Z}}(\frac{\pi}{2^l})$ gate is conjugated by a ladder of CNOTs along the support of the logical $Z$ operator to perform the operator $e^{\frac{-i\pi}{2^{l+1}}Z_0...Z_{d-1}} = R_{\overline{Z}}(\frac{\pi}{2^l})$. We can implement multi-controlled $R_Z(\nicefrac{\pi}{2^l})$ rotations by controlling the $R_Z(\nicefrac{\pi}{2^l})$ as in Fig.~\ref{fig:nonft arb}. To detect logical errors induced by this gate, we use a similar sequence of flag circuits as before, interleaving the rotations with gauge operators formed by propagating $\bar{X}$ operators through the multi-controlled gates. The first few steps of the sequence are illustrated in Fig.~\ref{fig:general ft gauge sequence}.

\begin{figure}
    \resizebox{\textwidth}{!}{\input{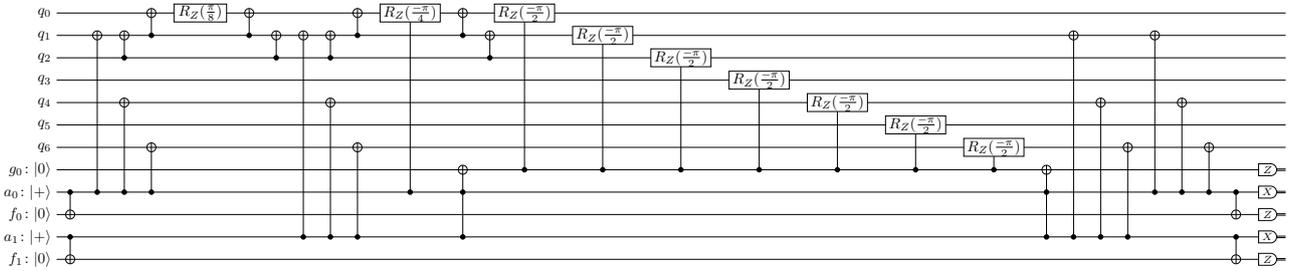}}
    \caption{Circuit for producing an $O(p^2)$-infidelity $\ket{\overline{\nicefrac{\pi}{8}}}$ state in the Steane code where $p$ is the physical error rate. This circuit takes as input a fault-tolerantly encoded $\ket{\overline{+}}$ state in the Steane code, which is then rotated to the $\ket{\overline{\nicefrac{\pi}{8}}}$ state by the circuit above. A logical operator on the Steane code caused by any single fault will be detected by the bottom four flag qubits or measurement of the garbage qubit. If any of the flag measurements are non-trivial, the state preparation should be aborted. A round of error detection must be performed on the state before injecting it into the Steane code.}
    \label{fig:Steane pi over 8}
\end{figure}

\begin{figure}
    \centering
    \includegraphics[width=0.45\linewidth]{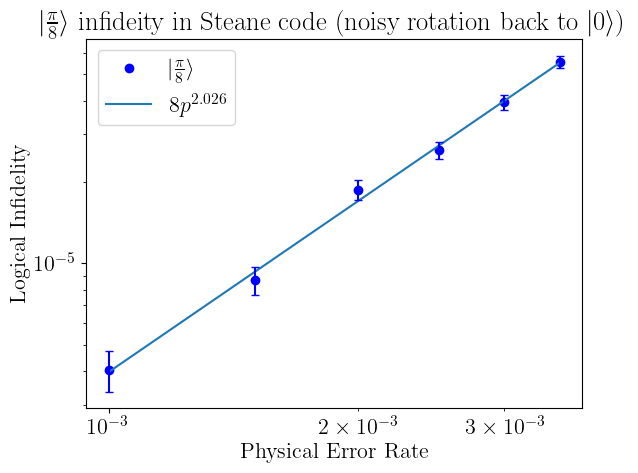} \includegraphics[width=0.45\linewidth]{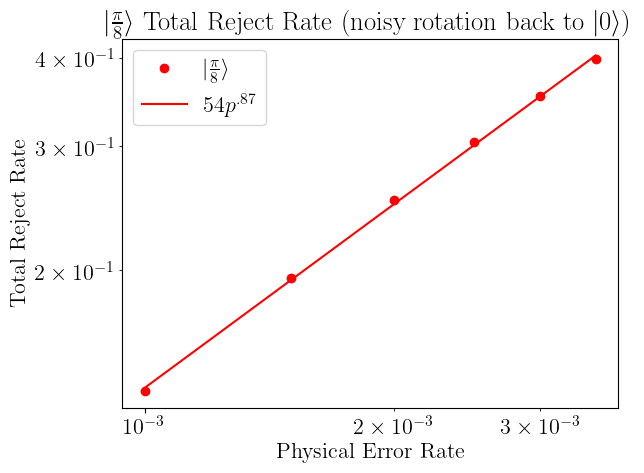}
    \caption{State-vector simulations varying the noise strength parameter $p$ for fault-tolerantly preparing a $\ket{\overline{\frac{\pi}{8}}}$ state in the Steane code with failure probability $O(p^2)$. Circuits were written in Guppy and state-vector simulations were performed in Selene as above.}
    \label{fig:pi 8 Steane sims}
\end{figure}

The recursion will terminate when the $R_Z(\nicefrac{\pi}{2^{l-m}})$ is transversal. This will certainly be the case for $R_Z(\pi)$, since the logical Z operator is always transversal, but it may occur earlier, as is the case with the Steane code where the $R_{\overline{Z}}(\nicefrac{\pi}{2})$ rotation is transversal. We now give our formula for a fault-distance-two $\overline{R_{\bar{Z}}(\frac{\pi}{2^l})}^{FT}$ circuit. Our fault-tolerant $\ket{\frac{\pi}{2^l}}$ state preparation circuit will simply be $\ket{\frac{\pi}{2^l}} = \overline{R_{\bar{Z}}(\frac{\pi}{2^l})}^{FT}\ket{+}$ where $\ket{+}$ is prepared fault-tolerantly using the method in \cite{Goto2016} and the $\ket{\frac{\pi}{2^l}}$ preparation circuit succeeds if no measurement returns a 1 in the $\overline{R_{\bar{Z}}(\frac{\pi}{2^l})}^{FT}$ circuit or the 
$\ket{+}$ state preparation circuit, and no non-trivial syndromes are recorded in a subsequent round of QED. Since the $R_{\overline{Z}}(\nicefrac{\pi}{2})$ rotation is transversal in the Steane code, our base case is 

\begin{equation}
        \centering
        \resizebox{\textwidth}{!}{\input{Figures/steane_base.tikz}}
\end{equation}
where we take as representatives of our logical $X$ and $Z$ operators $\overline{X} = X_1 X_4 X_6$ and $\overline{Z} = Z_0 Z_1 Z_2$.

and the recursive step is
        \begin{equation}
        \centering
        \resizebox{\textwidth}{!}{\begin{tikzpicture}[scale=1.000000,x=1pt,y=1pt]
\filldraw[color=white] (0.000000, -7.500000) rectangle (449.000000, 172.500000);
\draw[color=black] (0.000000,165.000000) -- (449.000000,165.000000);
\draw[color=black] (0.000000,165.000000) node[left] {$q_0$};
\draw[color=black] (0.000000,150.000000) -- (449.000000,150.000000);
\draw[color=black] (0.000000,150.000000) node[left] {$q_1$};
\draw[color=black] (0.000000,135.000000) -- (449.000000,135.000000);
\draw[color=black] (0.000000,135.000000) node[left] {$q_2$};
\draw[color=black] (0.000000,120.000000) -- (449.000000,120.000000);
\draw[color=black] (0.000000,120.000000) node[left] {$q_3$};
\draw[color=black] (0.000000,105.000000) -- (449.000000,105.000000);
\draw[color=black] (0.000000,105.000000) node[left] {$q_4$};
\draw[color=black] (0.000000,90.000000) -- (449.000000,90.000000);
\draw[color=black] (0.000000,90.000000) node[left] {$q_5$};
\draw[color=black] (0.000000,75.000000) -- (449.000000,75.000000);
\draw[color=black] (0.000000,75.000000) node[left] {$q_6$};
\draw[color=black] (0.000000,60.000000) -- (449.000000,60.000000);
\draw[color=black] (0.000000,60.000000) node[left] {$a_{0}$};
\draw[color=black] (0.000000,45.000000) node[anchor=mid east] {$\vdots$};
\draw[color=black] (0.000000,30.000000) -- (449.000000,30.000000);
\draw[color=black] (0.000000,30.000000) node[left] {$a_{m-1}$};
\draw[color=black] (113.000000,15.000000) -- (437.000000,15.000000);
\draw[color=black] (437.000000,14.500000) -- (449.000000,14.500000);
\draw[color=black] (437.000000,15.500000) -- (449.000000,15.500000);
\draw[color=black] (113.000000,0.000000) -- (437.000000,0.000000);
\draw[color=black] (437.000000,-0.500000) -- (449.000000,-0.500000);
\draw[color=black] (437.000000,0.500000) -- (449.000000,0.500000);
\draw (38.500000,165.000000) -- (38.500000,30.000000);
\begin{scope}
\draw[fill=white] (38.500000, 150.000000) +(-45.000000:45.961941pt and 29.698485pt) -- +(45.000000:45.961941pt and 29.698485pt) -- +(135.000000:45.961941pt and 29.698485pt) -- +(225.000000:45.961941pt and 29.698485pt) -- cycle;
\clip (38.500000, 150.000000) +(-45.000000:45.961941pt and 29.698485pt) -- +(45.000000:45.961941pt and 29.698485pt) -- +(135.000000:45.961941pt and 29.698485pt) -- +(225.000000:45.961941pt and 29.698485pt) -- cycle;
\draw (38.500000, 150.000000) node {$\overline{R_{\overline{Z}}(\pm \frac{\pi}{2^l})}^{FT}$};
\end{scope}
\filldraw (38.500000, 60.000000) circle(1.500000pt);
\filldraw (38.500000, 30.000000) circle(1.500000pt);
\draw[fill=white,color=white] (83.000000, -6.000000) rectangle (98.000000, 171.000000);
\draw (90.500000, 82.500000) node {$=$};
\filldraw[color=white] (110.000000, 12.000000) rectangle (116.000000, 18.000000);
\draw (116.000000, 12.000000) -- (116.000000, 18.000000);
\draw (113.000000, 15.000000) node {$\scriptstyle{\hspace{-30 pt}a_{m}\colon|+\rangle}$};
\filldraw[color=white] (110.000000, -3.000000) rectangle (116.000000, 3.000000);
\draw (116.000000, -3.000000) -- (116.000000, 3.000000);
\draw (113.000000, 0.000000) node {$\scriptstyle{\hspace{-30 pt}f_{m}\colon|0\rangle}$};
\draw (131.000000,15.000000) -- (131.000000,0.000000);
\begin{scope}
\draw[fill=white] (131.000000, 0.000000) circle(3.000000pt);
\clip (131.000000, 0.000000) circle(3.000000pt);
\draw (128.000000, 0.000000) -- (134.000000, 0.000000);
\draw (131.000000, -3.000000) -- (131.000000, 3.000000);
\end{scope}
\filldraw (131.000000, 15.000000) circle(1.500000pt);
\draw (149.000000,150.000000) -- (149.000000,15.000000);
\begin{scope}
\draw[fill=white] (149.000000, 150.000000) circle(3.000000pt);
\clip (149.000000, 150.000000) circle(3.000000pt);
\draw (146.000000, 150.000000) -- (152.000000, 150.000000);
\draw (149.000000, 147.000000) -- (149.000000, 153.000000);
\end{scope}
\filldraw (149.000000, 15.000000) circle(1.500000pt);
\draw (167.000000,105.000000) -- (167.000000,15.000000);
\begin{scope}
\draw[fill=white] (167.000000, 105.000000) circle(3.000000pt);
\clip (167.000000, 105.000000) circle(3.000000pt);
\draw (164.000000, 105.000000) -- (170.000000, 105.000000);
\draw (167.000000, 102.000000) -- (167.000000, 108.000000);
\end{scope}
\filldraw (167.000000, 15.000000) circle(1.500000pt);
\draw (185.000000,75.000000) -- (185.000000,15.000000);
\begin{scope}
\draw[fill=white] (185.000000, 75.000000) circle(3.000000pt);
\clip (185.000000, 75.000000) circle(3.000000pt);
\draw (182.000000, 75.000000) -- (188.000000, 75.000000);
\draw (185.000000, 72.000000) -- (185.000000, 78.000000);
\end{scope}
\filldraw (185.000000, 15.000000) circle(1.500000pt);
\draw (225.000000,165.000000) -- (225.000000,30.000000);
\begin{scope}
\draw[fill=white] (225.000000, 150.000000) +(-45.000000:35.355339pt and 29.698485pt) -- +(45.000000:35.355339pt and 29.698485pt) -- +(135.000000:35.355339pt and 29.698485pt) -- +(225.000000:35.355339pt and 29.698485pt) -- cycle;
\clip (225.000000, 150.000000) +(-45.000000:35.355339pt and 29.698485pt) -- +(45.000000:35.355339pt and 29.698485pt) -- +(135.000000:35.355339pt and 29.698485pt) -- +(225.000000:35.355339pt and 29.698485pt) -- cycle;
\draw (225.000000, 150.000000) node {${R_{\overline{Z}}(\pm \frac{\pi}{2^l})}$};
\end{scope}
\filldraw (225.000000, 60.000000) circle(1.500000pt);
\filldraw (225.000000, 30.000000) circle(1.500000pt);
\draw (304.500000,165.000000) -- (304.500000,15.000000);
\begin{scope}
\draw[fill=white] (304.500000, 150.000000) +(-45.000000:60.104076pt and 29.698485pt) -- +(45.000000:60.104076pt and 29.698485pt) -- +(135.000000:60.104076pt and 29.698485pt) -- +(225.000000:60.104076pt and 29.698485pt) -- cycle;
\clip (304.500000, 150.000000) +(-45.000000:60.104076pt and 29.698485pt) -- +(45.000000:60.104076pt and 29.698485pt) -- +(135.000000:60.104076pt and 29.698485pt) -- +(225.000000:60.104076pt and 29.698485pt) -- cycle;
\draw (304.500000, 150.000000) node {$\overline{R_{\overline{Z}}(\mp \frac{\pi}{2^l-1})}^{FT,ZZ}$};
\end{scope}
\filldraw (304.500000, 60.000000) circle(1.500000pt);
\filldraw (304.500000, 30.000000) circle(1.500000pt);
\filldraw (304.500000, 15.000000) circle(1.500000pt);
\draw (362.000000,150.000000) -- (362.000000,15.000000);
\begin{scope}
\draw[fill=white] (362.000000, 150.000000) circle(3.000000pt);
\clip (362.000000, 150.000000) circle(3.000000pt);
\draw (359.000000, 150.000000) -- (365.000000, 150.000000);
\draw (362.000000, 147.000000) -- (362.000000, 153.000000);
\end{scope}
\filldraw (362.000000, 15.000000) circle(1.500000pt);
\draw (380.000000,105.000000) -- (380.000000,15.000000);
\begin{scope}
\draw[fill=white] (380.000000, 105.000000) circle(3.000000pt);
\clip (380.000000, 105.000000) circle(3.000000pt);
\draw (377.000000, 105.000000) -- (383.000000, 105.000000);
\draw (380.000000, 102.000000) -- (380.000000, 108.000000);
\end{scope}
\filldraw (380.000000, 15.000000) circle(1.500000pt);
\draw (398.000000,75.000000) -- (398.000000,15.000000);
\begin{scope}
\draw[fill=white] (398.000000, 75.000000) circle(3.000000pt);
\clip (398.000000, 75.000000) circle(3.000000pt);
\draw (395.000000, 75.000000) -- (401.000000, 75.000000);
\draw (398.000000, 72.000000) -- (398.000000, 78.000000);
\end{scope}
\filldraw (398.000000, 15.000000) circle(1.500000pt);
\draw (416.000000,15.000000) -- (416.000000,0.000000);
\begin{scope}
\draw[fill=white] (416.000000, 0.000000) circle(3.000000pt);
\clip (416.000000, 0.000000) circle(3.000000pt);
\draw (413.000000, 0.000000) -- (419.000000, 0.000000);
\draw (416.000000, -3.000000) -- (416.000000, 3.000000);
\end{scope}
\filldraw (416.000000, 15.000000) circle(1.500000pt);
\draw[fill=white] (431.000000, 11.000000) -- (439.000000,11.000000) arc (-90:90:4.000000pt) -- (431.000000,19.000000) -- cycle;
\draw (437.000000, 15.000000) node {{\scriptsize $X$}};
\draw[fill=white] (431.000000, -4.000000) -- (439.000000,-4.000000) arc (-90:90:4.000000pt) -- (431.000000,4.000000) -- cycle;
\draw (437.000000, 0.000000) node {{\scriptsize $Z$}};
\draw[color=black] (449.000000,45.000000) node[anchor=mid west] {$\vdots$};
\end{tikzpicture}}
    \end{equation}
The $\ket{\overline{\nicefrac{\pi}{8}}}$ circuit with failure probability $O(p^2)$ in the Steane code is shown in Fig.~\ref{fig:Steane pi over 8}. (Note that we do not require a flag $a_z$ to check for logical $X$ errors since a single failure in the circuit cannot produce that error, in contrast to the $d=2$ case). State-vector simulations with a depolarizing error model showing an $O(p^2)$ scaling rate are given in Fig.~\ref{fig:pi 8 Steane sims}. 

\section{Higher fault-distance Clifford circuits}
\begin{figure}
    \resizebox{\textwidth}{!}{\input{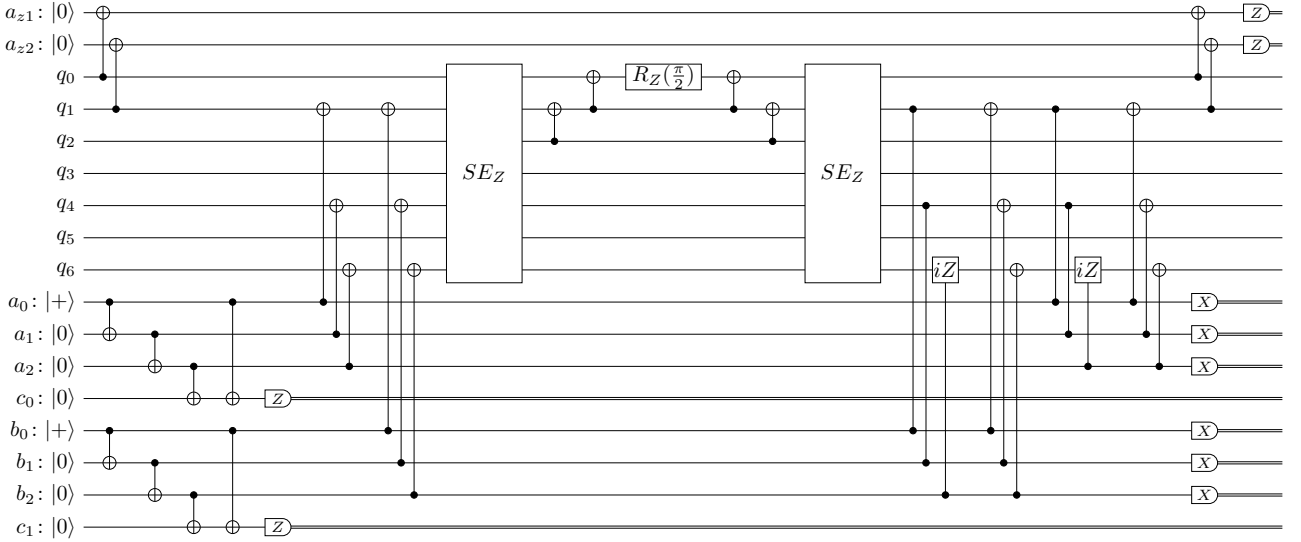}}
    \caption{Fault distance 3 $R_{\overline{Z}}(\pi/2)$ circuit in the Steane code. We implement the $R_{\overline{Z}}(\pi/2)$ non-transversally as we would a $T$ gate. Two cat states are initialized and then gauge operators of the $R_{\overline{Z}}(\pi/2)$ gate are stored in them. The $iZZZ$ terms of these operators would be replaced by a transversal $R_{\overline{Z}}(\pi/2)$ gate in the $R_{\overline{Z}}(\pi/4)$ version of the circuit, and the cat states would be enlarged to seven qubits each. The top flags $a_{z1}$ and $a_{z2}$ are added to ensure a fault-distance of 3 to $X$ errors as well. The $SE_Z$ boxes refer to fault-tolerant extraction of the Steane code $Z$ syndromes.}
    \label{fig:Steane pi over 2}
\end{figure}

First, we give a fault-distance 3 $R_{\overline{Z}}(\nicefrac{\pi}{2})$ gate in the Steane code, which we implement non-transversally to mimic the structure of the $T$ gate. To increase the fault-distance of our circuit to 3, we need to repeat the measurement of the gauge operator checking for logical $Z$ errors, as well as to correct $X$ errors which may also induce a logical $Z$ error by performing one half of a QEC cycle, which we perform via Steane-style error correction \cite{Steane_1997}, before and after the $R_{\overline{Z}}(\nicefrac{\pi}{2})$ gate but after we have begun the measurement of the gauge operators. We also use a three-qubit cat state instead of Bell-pair ancilla to differentiate between hook errors. Altogether, we obtain the circuit in Fig.~\ref{fig:Steane pi over 2}
Using \texttt{Stim}'s\cite{Gidney_2021} minimum-fault-finding function, we verify that this circuit has a fault distance of 3. 

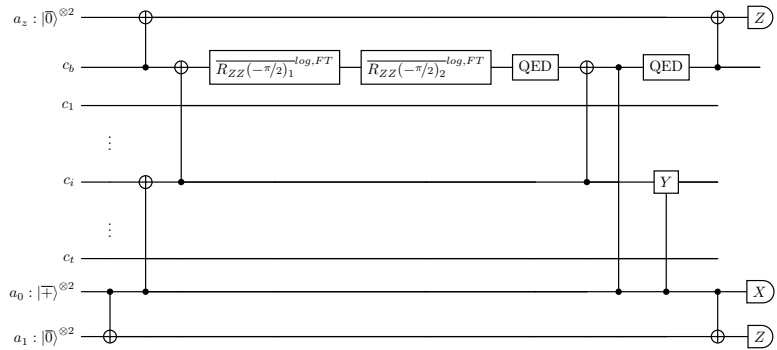
\begin{figure}[htpb!]
    \centering
    \resizebox{.6\textwidth}{!}{
    \begin{quantikz}
    \lstick{$a_{z}: \ket{\overline{0}}^{\otimes 2}$} &&\targ{}&\qw&\qw&\qw&\qw&\qw&\qw&\qw&\targ{}&\meterD{Z}  \\
    \lstick{$c_b$} &&\ctrl{-1}&\targ{}&\gate{\overline{{R_{ZZ}}(-\nicefrac{\pi}{2})_1}^{log, FT}}&\gate{\overline{{R_{ZZ}}(-\nicefrac{\pi}{2})_2}^{log, FT}}&\gate{\text{QED}}&\targ{}&\ctrl{0} &\gate{\text{QED}}&\ctrl{-1}&\qw  \\
    \lstick{$c_1$} && & &\qw&\qw&\qw&\qw&\qw&\qw&\qw  \\
    &\setwiretype{n} \vdots &&  & & &  \\ 
    \lstick{$c_i$} &&\targ{}&\ctrl{-3}&\qw&\qw&\qw&\ctrl{-3}&\qw&\gate{Y}&\qw  \\
    &\setwiretype{n} \vdots & & & &  \\ 
    \lstick{$c_{t}$} & & &\qw&\qw&\qw&\qw&\qw&\qw&\qw&\qw  \\
    \lstick{$a_0: \ket{\overline{+}}^{\otimes 2}$} &\ctrl{1}&\ctrl{-3}&\qw&\qw&\qw&\qw&\qw&\ctrl{-6}&\ctrl{-3}&\ctrl{1}&\meterD{X}  \\
    \lstick{$a_1: \ket{\overline{0}}^{\otimes 2}$} &\targ{}&\qw&\qw&\qw&\qw&\qw&\qw&\qw&\qw&\targ{}&\meterD{Z}  \\
    \end{quantikz}}
    \caption{Concatenated logical $R_Z(\nicefrac{\pi}{2})
    \otimes R_Z(\nicefrac{\pi}{2})$ circuit on two logical qubits in the $\nkd{16}{4}{4}$ code using logical $\nkd{4}{2}{2}$ ancillae $a_z,a_0,$ and $a_1$ and the $\overline{{R_{ZZ}}(-\nicefrac{\pi}{2})_j}^{log,FT}$ for $d=2$ on the bottom $\nkd{4}{2}{2}$ block, $c_b$ of the $\nkd{16}{4}{4}$ code. The controlled-$Y$ gate is performed fault-tolerantly to distance $2$ by conjugating a transversal CX gate with the fault-tolerant $\overline{S}^{\otimes 2}$ gadget of \cite{Chao_2018}, performing a $Z$ syndrome extraction after each $\overline{S}^{\otimes 2}$ or $\overline{S^\dagger}^{\otimes 2}$ gadget. The CZ gate between $a_0$ and $c_b$ is transversal, but since to perform this logical CZ on both qubits, the top qubit of one block must be paired with the bottom qubit of the other, the $CZ$ does not have the same qubit pairing as with the other gates. Therefore another round of full $d=2$ syndrome extraction (denoted by QED) must be performed on $c_b$ after the gate. }\label{fig:concatenated RZpi2}
\end{figure}

Another possible low-overhead approach to obtaining higher-distance versions of the $d=2$ iceberg $\overline{{R_{\overline{Z}}}(\nicefrac{\pi}{2^l})}^{log,FT}$ would be to perform the same circuits at the logical level, with all the gadgets replaced by their fault-distance-two counterparts. The flag qubits would become $d=2$ ancillae and the CX gates would become transversal CX gates. Physical $R_Z(\nicefrac{\pi}{2^l})$ gates would be replaced with the distance-two $\overline{{R_{\overline{Z}}}(\nicefrac{\pi}{2^l})}^{FT}$ gates from Section \ref{sec:d=2 iceberg}, followed by $d=2$ syndrome extraction. Although these non-Clifford circuits, even for the smallest concatenated iceberg code (with parameters $\nkd{16}{4}{4}$), would require enough qubits to make direct state-vector simulation difficult when taking into account the ancilla overhead, we can investigate concatenating the $\overline{{R_{ZZ}}_{q_i,q_b}(-\nicefrac{\pi}{2})}^{FT}$ circuit with itself, since this is a Clifford circuit amenable to stabilizer simulation. Again, using \texttt{Stim}'s minimum-fault-finding function, we verify that the circuit shown in Fig.~\ref{fig:concatenated RZpi2} has a fault distance of 4. We test fault-distance of the gate by noiselessly preparing either the $|\overline{Y^+Y^+00}\rangle$ or $|\overline{0000}\rangle$ state, performing the circuit in Fig.~\ref{fig:concatenated RZpi2}, and then performing with noise the half of the QEC cycle introduced in \cite{dasu2026} for the $[[16,4,4]]$ code which extracts $Z$ syndromes before measuring in the appropriate basis. Again, using \texttt{Stim}'s minimum-fault-finding function, we verify that the overall circuits have a fault distance of 4.

\section{Summary and Outlook}

We have introduced a sequence of gauge operators that will detect logical errors induced by non-fault-tolerant $R_{\overline{Z}}(\nicefrac{\pi}{2^l})$ rotations on CSS codes. In addition to it being of theoretical interest that the logical errors induced by such $\nicefrac{\pi}{2^l}$ rotations can, in fact, be flagged, we believe that these circuits are also of practical value. The fidelities achieved for the iceberg code for the $T$ and $\sqrt{T}$ gates at $p = 10^{-3}$ or $p = 10^{-4}$ are beyond what one would expect from the partially fault-tolerant technique of simply applying the non-fault tolerant $e^{iZ_bZ_1 \frac{\pi}{2^l}}$ two-qubit gate as in \cite{Self_2024} and post-selecting on non-trivial stabilizers, since it would be precisely the $XX$, $YY$ and $ZZ$ terms that the flagging apparatus introduced above detects in the fully-fault-tolerant case that would cause errors in the non-fault-tolerant case. Using a two-qubit depolarizing error model with strength $p$ for the $e^{iZ_bZ_1 \frac{\pi}{2^l}}$ gate, these occur with probability $\nicefrac{p}{5}$ which is greater at realistic noise than for the fully fault-tolerant scaling of $\sim 10 p^2$ observed for our $T$ and $\sqrt{T}$ circuits. That being said, in concatenated codes, this technique can also be combined with partially fault-tolerant techniques. For instance, lower-level non-Cliffords can be performed with low-overheads using partially-fault-tolerant techniques and then flagged using this construction. Also, the fault-tolerant preparation of $\ket{\nicefrac{\pi}{2^l}}$ states in distance-three codes could be useful in performing algorithmic primitives on near-term devices as well as in improving synthesis overheads through allowing for at least Clifford+$\sqrt{T}$ synthesis \cite{Kliuchnikov_2023}. 

Although we do not investigate distances beyond four in this work due to difficulties of simulation, generalizing these techniques to higher distances is a promising direction for future research. For instance, one could attempt to generalize cultivation~\cite{gidney2024} to $\ket{\nicefrac{\pi}{2^l}}$ states in the Steane code or higher-distance color codes based on measurements of these gauge operators as illustrated in Fig.~\ref{fig:Steane pi over 2}. An alternative direction of research would be to concatenate the $d=2$ iceberg circuits with themselves to obtain higher-distance fault-tolerant $R_{\overline{Z}}(\nicefrac{\pi}{2^l})$ gates, similar to the circuit concatenations in \cite{stepbystep, dasu2025} for magic state preparation. Although we have demonstrated that this approach works for the $R_{\overline{Z}}(\nicefrac{\pi}{2})$ gate, assessing the fault tolerance of the concatenation of the non-Clifford circuits for higher $l$ is outside the reach of direct state-vector simulation and will require different tools for analysis.

\section*{Acknowledgments}
The authors would like to thank Ali Lavasani, Callum Macpherson, Karl Mayer, Noah Berthusen, David Stephen, David Amaro, Serban Cercelescu, Mark Koch, Tyler LeBlond, Drew Potter, Matthew DeCross, Pablo Andres-Martinez, and David Hayes for helpful and fun conversations.

\section*{Author Contributions}
S.D. and B.C. independently conceived of the idea of measuring non-Pauli gauge operators of non-Clifford rotations on the iceberg code. S.D. conceived of the idea to use a recursive sequence of gauge operators to perform $\nicefrac{\pi}{2^l}$ gates and the circuits to do so. B.C. improved the implementation of the controlled-rotation operators. S.D. conceived of the amortized linear gate-count and depth multi-controlled operator implementation. Both authors contributed to the writing and editing of the manuscript.
\clearpage
\printbibliography 

@article{Chao_2018,
   title={Fault-tolerant quantum computation with few qubits},
   volume={4},
   ISSN={2056-6387},
   url={http://dx.doi.org/10.1038/s41534-018-0085-z},
   DOI={10.1038/s41534-018-0085-z},
   number={1},
   journal={npj Quantum Information},
   publisher={Springer Science and Business Media LLC},
   author={Chao, Rui and Reichardt, Ben W.},
   year={2018},
   month=sep }

@misc{delfosse2023,
      title={Spacetime codes of Clifford circuits}, 
      author={Nicolas Delfosse and Adam Paetznick},
      year={2023},
      eprint={2304.05943},
      archivePrefix={arXiv},
      primaryClass={quant-ph},
      url={https://arxiv.org/abs/2304.05943}, 
}

@article{Goto2016,
	author = {Goto, Hayato},
	date = {2016/01/27},
	doi = {10.1038/srep19578},
	id = {Goto2016},
	isbn = {2045-2322},
	journal = {Scientific Reports},
	number = {1},
	pages = {19578},
	title = {Minimizing resource overheads for fault-tolerant preparation of encoded states of the Steane code},
	url = {https://doi.org/10.1038/srep19578},
	volume = {6},
	year = {2016}}

@misc{koch2025,
      title={GUPPY: Pythonic Quantum-Classical Programming}, 
      author={Mark Koch and Alan Lawrence and Kartik Singhal and Seyon Sivarajah and Ross Duncan},
      year={2025},
      eprint={2510.12582},
      archivePrefix={arXiv},
      primaryClass={cs.PL},
      url={https://arxiv.org/abs/2510.12582}, 
}

@misc{Selene,
  author = {Quantinuum},
  title = {Selene},
  year = {2025},
  url = {https://github.com/Quantinuum/selene},
}

@article{Self_2024,
   title={Protecting expressive circuits with a quantum error detection code},
   volume={20},
   ISSN={1745-2481},
   url={http://dx.doi.org/10.1038/s41567-023-02282-2},
   DOI={10.1038/s41567-023-02282-2},
   number={2},
   journal={Nature Physics},
   publisher={Springer Science and Business Media LLC},
   author={Self, Chris N. and Benedetti, Marcello and Amaro, David},
   year={2024},
   month=jan, pages={219–224} }

@misc{gidney2024,
      title={Magic state cultivation: growing T states as cheap as CNOT gates}, 
      author={Craig Gidney and Noah Shutty and Cody Jones},
      year={2024},
      eprint={2409.17595},
      archivePrefix={arXiv},
      primaryClass={quant-ph},
      url={https://arxiv.org/abs/2409.17595}, 
}

@article{stepbystep,
author = {Goto, Hayato},
year = {2014},
month = {12},
pages = {7501},
title = {Step-by-step magic state encoding for efficient fault-tolerant quantum computation},
volume = {4},
journal = {Scientific reports},
doi = {10.1038/srep07501}
}

@misc{dasu2025,
      title={Breaking even with magic: demonstration of a high-fidelity logical non-Clifford gate}, 
      author={Shival Dasu and Simon Burton and Karl Mayer and David Amaro and Justin A. Gerber and Kevin Gilmore and Dan Gresh and Davide DelVento and Andrew C. Potter and David Hayes},
      year={2025},
      eprint={2506.14688},
      archivePrefix={arXiv},
      primaryClass={quant-ph},
      url={https://arxiv.org/abs/2506.14688}, 
}

@misc{xu2026controlledjumpcliffordhierarchy,
      title={Controlled jump in the Clifford hierarchy}, 
      author={Yichen Xu and Xiao Wang},
      year={2026},
      eprint={2602.22201},
      archivePrefix={arXiv},
      primaryClass={quant-ph},
      url={https://arxiv.org/abs/2602.22201}, 
}

@phdthesis{gottesmanPHD,
    author={Daniel Gottesman},
    title={{Stabilizer Codes and Quantum Error Correction}},
    school={California Institute of Technology},
    year={1997},
    doi={10.7907/rzr7-dt72},
    eprint={quant-ph/9705052},
    archivePrefix={arXiv},
}

@article{Gidney_2021,
   title={Stim: a fast stabilizer circuit simulator},
   volume={5},
   ISSN={2521-327X},
   url={http://dx.doi.org/10.22331/q-2021-07-06-497},
   DOI={10.22331/q-2021-07-06-497},
   journal={Quantum},
   publisher={Verein zur Forderung des Open Access Publizierens in den Quantenwissenschaften},
   author={Gidney, Craig},
   year={2021},
   month=jul, pages={497} }

@unpublished{CrigerInPreparation,
  author = {Ben Criger and others},
  title = {Automated Flag-based Fault-Tolerant State Preparation using Integer Linear Programming},
  year = {2026},
  note = {In preparation}
}

@misc{ross2016,
      title={Optimal ancilla-free Clifford+T approximation of z-rotations}, 
      author={Neil J. Ross and Peter Selinger},
      year={2016},
      eprint={1403.2975},
      archivePrefix={arXiv},
      primaryClass={quant-ph},
      url={https://arxiv.org/abs/1403.2975}, 
}

@inproceedings{Kitaev2002,
  title={Classical and Quantum Computation},
  author={Alexei Y. Kitaev and Ao Shen and Mikhail N. Vyalyi},
  booktitle={Graduate Studies in Mathematics},
  year={2002},
  url={https://api.semanticscholar.org/CorpusID:265878561}
}

@article{BaconFlammiaHarrowShi,
  author={Bacon, Dave and Flammia, Steven T. and Harrow, Aram W. and Shi, Jonathan},
  journal={IEEE Transactions on Information Theory}, 
  title={Sparse Quantum Codes From Quantum Circuits}, 
  year={2017},
  volume={63},
  number={4},
  pages={2464-2479},
  doi={10.1109/TIT.2017.2663199},
  ISSN={1557-9654},
  month={April},}

@misc{kim2026,
      title={Catalytic $z$-rotations in constant $T$-depth}, 
      author={Isaac H. Kim},
      year={2026},
      eprint={2506.15147},
      archivePrefix={arXiv},
      primaryClass={quant-ph},
      url={https://arxiv.org/abs/2506.15147}, 
}

@article{Gidney_2018,
   title={Halving the cost of quantum addition},
   volume={2},
   ISSN={2521-327X},
   url={http://dx.doi.org/10.22331/q-2018-06-18-74},
   DOI={10.22331/q-2018-06-18-74},
   journal={Quantum},
   publisher={Verein zur Forderung des Open Access Publizierens in den Quantenwissenschaften},
   author={Gidney, Craig},
   year={2018},
   month=jun, pages={74} }

@article{Berthusen_2025,
   title={Adaptive Syndrome Extraction},
   volume={6},
   ISSN={2691-3399},
   url={http://dx.doi.org/10.1103/ps3r-wf84},
   DOI={10.1103/ps3r-wf84},
   number={3},
   journal={PRX Quantum},
   publisher={American Physical Society (APS)},
   author={Berthusen, Noah and Tan, Shi Jie Samuel and Huang, Eric and Gottesman, Daniel},
   year={2025},
   month=jul }

@article{Kliuchnikov_2023,
   title={Shorter quantum circuits via single-qubit gate approximation},
   volume={7},
   ISSN={2521-327X},
   url={http://dx.doi.org/10.22331/q-2023-12-18-1208},
   DOI={10.22331/q-2023-12-18-1208},
   journal={Quantum},
   publisher={Verein zur Forderung des Open Access Publizierens in den Quantenwissenschaften},
   author={Kliuchnikov, Vadym and Lauter, Kristin and Minko, Romy and Paetznick, Adam and Petit, Christophe},
   year={2023},
   month=dec, pages={1208} }

@article{Steane_1997,
   title={Active Stabilization, Quantum Computation, and Quantum State Synthesis},
   volume={78},
   ISSN={1079-7114},
   url={http://dx.doi.org/10.1103/PhysRevLett.78.2252},
   DOI={10.1103/physrevlett.78.2252},
   number={11},
   journal={Physical Review Letters},
   publisher={American Physical Society (APS)},
   author={Steane, A. M.},
   year={1997},
   month=mar, pages={2252–2255} }

@misc{dasu2026,
      title={Computing with many encoded logical qubits beyond break-even}, 
      author={Shival Dasu and Matthew DeCross and Andrew Y. Guo and Ali Lavasani and Jan Behrends and Asmae Benhemou and Yi-Hsiang Chen and Karl Mayer and Chris N. Self and Selwyn Simsek and Basudha Srivastava and M. S. Allman and Jake Arkinstall and Justin G. Bohnet and Nathaniel Q. Burdick and J. P. Campora III and Alex Chernoguzov and Samuel F. Cooper and Robert D. Delaney and Joan M. Dreiling and Brian Estey and Caroline Figgatt and Cameron Foltz and John P. Gaebler and Alex Hall and Craig A. Holliman and Ali A. Husain and Akhil Isanaka and Colin J. Kennedy and Yuga Kodama and Nikhil Kotibhaskar and Nathan K. Lysne and Ivaylo S. Madjarov and Michael Mills and Alistair R. Milne and Brian Neyenhuis and Annie J. Park and Anthony Ransford and Adam P. Reed and Steven J. Sanders and Charles H. Baldwin and David Hayes and Ben Criger and Andrew C. Potter and David Amaro},
      year={2026},
      eprint={2602.22211},
      archivePrefix={arXiv},
      primaryClass={quant-ph},
      url={https://arxiv.org/abs/2602.22211}, 
}
\end{document}